\documentclass[aps,twocolumn,amsmath,amssymb,preprintnumbers]{revtex4}
\usepackage{amsmath} \usepackage{amsfonts} \usepackage{amssymb}
\usepackage{bbm}
\usepackage{epsfig}
\usepackage{graphics}
\usepackage{graphicx}
\newcommand{\be}{\begin{equation}}
\newcommand{\ee}{\end{equation}}
\newcommand{\ba}{\begin{eqnarray}}
\newcommand{\ea}{\end{eqnarray}}
\newcommand{\nn}{\nonumber}

\begin{document}

\title[ ]{Occupation numbers from functional integral}

\author{C. Wetterich}
\affiliation{Institut  f\"ur Theoretische Physik\\
Universit\"at Heidelberg\\
Philosophenweg 16, D-69120 Heidelberg}

\begin{abstract}
Occupation numbers for non-relativistic interacting particles are discussed within a functional integral formulation. We concentrate on zero temperature, where the Bogoliubov theory breaks down for strong couplings as well as for low dimensional models. We find that the leading behavior of the occupation numbers for small momentum is governed by a quadratic time derivative in the inverse propagator that is not contained in the Bogoliubov theory. We propose to use a functional renormalization group equation for the occupation numbers in order to implement systematic non-perturbative extensions beyond the Bogoliubov theory.
\end{abstract}

\maketitle

\section{Introduction}
Occupation numbers are the central tool for the statistical description of non-interacting particles or quasi-particles. Extending this concept to interacting particles allows for the description of many interesting phenomena, as condensates, the formation of gaps and pseudogaps or nontrivial scaling for the momentum dependence of the occupation numbers. Most of these phenomena occur at low temperatures T and we will concentrate on this case.

For non-relativistic bosons and small couplings the use of perturbation theory, within a Hamiltonian framework with operator valued fields, has led to the successful Bogoliubov theory \cite{B}. This theory breaks down, however, in low dimensions and for strong couplings. In one dimension $(d=1)$ and for $T=0$, the occupation number $n(\vec q)$ for a mode with momentum $\vec q$ diverges proportional to the inverse momentum, $n(\vec q)\sim q^{-1}, q=|\vec q|$. The particle density is the momentum integral over the occupation numbers and diverges logarithmically in this case. Furthermore, the Bogoliubov theory is built on a nonzero condensate density or order parameter $\bar\phi_0$. Such an order parameter leads to spontaneous symmetry breaking of the global $U(1)$-symmetry associated to the conserved particle number. It is known that for $d=1$ and $T=0$ no spontaneous symmetry breaking of a continuous symmetry is possible - the order parameter $\bar\phi_0$ has to vanish in the infinite volume limit. Also for $d=2$ the situation is problematic. Now one finds a nonzero $\bar\phi_0$ for $T=0$ and the Bogoliubov approximation to the particle density shows no infrared divergence anymore. However, for any nonzero temperature the order parameter must again vanish due to the Mermin-Wagner theorem \cite{MW}, $\bar\phi_0(T>0)=0$. This reveals again a problem for the perturbative treatment where $n(\vec q)\sim \bar\phi_0/q$ for small $q$. For $q>0$ the occupation numbers should be continuous functions of $T$ and this indicates a contradiction between the Bogoliubov approximation and the Mermin-Wagner theorem.

Going beyond leading order perturbation theory (Bogoliubov approximation) becomes quite intricate in the operator formalism. A good alternative is the functional integral formulation which has already given a rather detailed picture of Bose-Einstein condensation for interacting bosons, largely by use of the renormalization group \cite{RGBE}. In this paper we address the computation of occupation numbers within a functional integral approach.

Already at the very basic level of the definition of $n(\vec q)$ one needs to be careful to subtract ``counterterms'' that diverge in the ultraviolet limit. We work with a momentum cutoff $\vec q^2<\Lambda^2$ and show that a too naive definition of $n(\vec q)$ would lead to a divergent particle density for $\Lambda\to\infty$. One needs to introduce counterterms that arise from the transition from the Hamiltonian formalism to the functional integral. With this correct starting point the functional integral formalism is well suited to solve the infrared problems of the Bogoliubov theory, as well as potential ultraviolet problems of extensions of it.

For $T=0$ we show how the infrared problems of the low dimensional Bogoliubov theory can be resolved by an investigation of the frequency and momentum dependence of the full propagator. In particular, it is crucial that the frequency dependence of the inverse propagator is quadratic in the range of small $q$ \cite{CWQP}, and not linear as for the Bogoliubov theory. The use of the propagator computed in \cite{CWQP} removes the infrared problems. For $q\to 0$ the particle number behaves now $\sim V^{-1/2} q^{-1+\eta}$ instead of $\sim \bar\phi_0q^{-1}$ in the Bogoliubov approximation. Here $V$ is the coefficient of the term $\sim \omega^2$ in the inverse propagator, with $\omega$ the frequency. The anomalous dimension $\eta$ vanishes for $d=2,3$, while it is in the range $0<\eta\leq 1/2$ for $d=1$. This renders the particle number finite for all dimensions. Furthermore, the leading behavior of $n(\vec q)$ for $q\to 0$ becomes independent of $\bar\phi_0$ and is therefore not affected by the vanishing of $\bar\phi_0$ at $T=0$ (for $d=1$) or $T>0$ (for $d=2$). In three dimensions and for a small coupling $\lambda_\phi$ the behavior $n\sim V^{-1/2}q^{-1}$ is relevant only for very small $q$, whereas the validity of the Bogoliubov approximation $n\sim \lambda^{1/2}_\phi\bar\phi_0/q$ extends to a large range of $q$. 

Extensions beyond the Bogoliubov theory have also to cope with ultraviolet problems. They occur if the frequency dependence of the propagator is not taken into account properly. We propose here a renormalization group treatment based on the average action \cite{CWAV,CWFE,BTW}. An infrared cutoff $k$ for the fluctuations is introduced in order to extrapolate between the microscopic action for $k=\Lambda$ and the full quantum effective action for $k=0$. Correspondingly, we study flow equations for $k$-dependent occupation numbers $n_k(\vec q)$. They yield the physical occupation numbers $n(\vec q)$ as the IR-cutoff is removed for $k=0$. The renormalization group flow permits an improvement in the computation of $n(\vec q)$ since for every momentum range only the effective renormalized couplings in this momentum range matter. In particular, we use the renormalization group flow in order to show two properties. The first is the non-renormalization of the condensate density for non-relativistic bosons. It is given by the unrenormalized order parameter, $n_c=\bar\phi^2_0$. Second, this method is free of ultraviolet problems. Our methods can be extended to coupled systems of bosons and fermions (cf. sect. \ref{gapsandpseudo} and appendix B) which are relevant for the BEC-BCS crossover for ultracold fermions. There the issue of the condensate density is more complicated \cite{DW}. A renormalization group flow for the total density in such systems has already been investigated in \cite{DGPW}. 

This paper is organized as follows. In sect. \ref{occupationnumbers} we discuss the occupation numbers in a functional integral context. They are related to a $q_0$-integral over the propagator, with $q_0$ the Euclidean frequency (Matsubara frequency for $T>0$). This section also introduces the (ultraviolet divergent) counterterms. Sect. \ref{low} specializes to zero temperature and sect. \ref{Bosonoccupationnumbers} deals with the case where the frequency dependence of the inverse propagator is linear. This yields the Bogoliubov approximation and first extensions of it. In sect. \ref{dynamicterm} we extend the general form of the inverse propagator by taking terms quadratic in the frequency (or also higher orders) into account. We show that this solves the infrared problems of the Bogoliubov theory in low dimensions. We present differential equations for the momentum dependence of the occupation numbers and the particle density. In sect. \ref{improvement} we turn to a heuristic renormalization group improvement. This helps for a proper understanding of the occupation numbers for large momentum.

The following sections introduce an exact renormalization group equation for the occupation numbers. It motivates the simpler heuristic renormalization group improvements. We start in sect. \ref{sourceterms} by introducing a generating functional for the occupation numbers. This can be made scale dependent, and an exact flow equation is derived in sect. \ref{flow}. We perform a non-perturbative approximation by quadratic truncation in sect. \ref{quadratic}.The flow shows a mutual influence of occupation numbers for different momenta - it cannot be separately described for individual momentum modes. The corresponding density transfer function is discussed in sect. \ref{density}. In sect. \ref{densityflow} we apply the functional renormalization group equations to non-relativistic bosons for $T=0$. Sect. \ref{gapsandpseudo} discusses gaps and pseudogaps in the light of renormalization group improvement and we give a short summary of our main results in sect. \ref{conclusions}.

\vspace{-0.5cm}
\section{Occupation numbers and renormalization}
\label{occupationnumbers}
We start with the partition function for a nonrelativistic bosonic particle
\begin{eqnarray}\label{1}
Z&=&\int {\cal D}\chi\exp(-S[\chi]),\nonumber\\
S[\chi]&=&\int_x(\chi^*\partial_\tau\chi+\frac{1}{2M}\vec\nabla\chi^*\vec\nabla\chi-\sigma\chi^*\chi)+S_{int}.
\end{eqnarray}
The complex field $\chi$ may be expressed by its Fourier modes
\begin{equation}\label{2}
\chi(x)=\chi(\tau,\vec x)=\int\limits_{\vec q}e^{i\vec q\vec x}\chi(\tau,\vec q)=\int_qe^{iqx}\chi(q),
\end{equation}
with
\ba 
q =(q_0,\vec q)~,~\int_{\vec q}&=&(2\pi)^{-d}\int d^d\vec q~,\nonumber\\
~\int_q&=&(2\pi)^{(d+1)}\int dq_0\int d^d\vec q.
\ea
For nonzero temperature $T$ the Euclidean time $\tau$ parameterizes a circle with circumference $\Omega_\tau=T^{-1}$ and the Matsubara frequencies $q_0=2\pi nT~,~n\in{\mathbbm Z}$, are discrete, with $\int_{q_0}=T\sum\limits_n$. The second thermodynamic parameter is the effective chemical potential $\sigma$. We regularize the theory by a momentum cutoff $\vec q\ ^2<\Lambda^2$ and take $\Lambda\to\infty$ when appropriate. 

Furthermore, we assume the invariance of the classical action $S$ under a global abelian symmetry of phase rotations  $\chi\to e^{i\phi}\chi$, corresponding to a conserved total particle number
\begin{equation}\label{3}
N=\int_{\vec x}n(\vec x)=\Omega_d\int_{\vec q}n(\vec q),
\end{equation}
with $\Omega_d$ the volume of $d$-dimensional space $(\Omega_{d+1}=\Omega_d\Omega_\tau)$. In our normalization $n(\vec q)$ is dimensionless. The momentum distribution $n(\vec q)$ denotes the fraction of particles with momentum $\vec q$. This is the quantity we want to investigate in this paper. For definiteness, we may consider a torus with finite $\Omega_d$ and discrete momenta, taking $\Omega_d\to\infty$ at the end. For discrete momenta we write 
\begin{equation}\label{3A}
N=\sum_{\vec q} n(\vec q)
\end{equation}
and identify $n(\vec{q})$ with the occupation number in momentum space. 

Following the Noether construction we can express $n(\vec x)$ as
\begin{equation}\label{3A1}
n(\vec x)=\frac{1}{\Omega_\tau}\int d\tau\langle\chi^*(\tau,\vec x)\chi(\tau,\vec x)\rangle-\frac12\int
\frac{d^d\vec q}{(2\pi)^3}
\end{equation}
and, correspondingly 
\begin{equation}\label{4}
n(\vec{q})=\frac{1}{\Omega_{d+1}}\int_{q_0}\langle\chi^*(q_0,\vec q)\chi(q_0,\vec q)\rangle-\frac12.
\end{equation}
Note that $n(\vec q)$ is not the Fourier transform of $n(\vec x)$. 

The constant part $-1/2$ in $n(\vec{q})$ arises from the connection between the functional integral and the operator formalism \cite{DW}. For a single degree of freedom it is the symmetric combination of annihilation and creation operators $(a^\dagger a+aa^\dagger)/2$ that translates into $\chi^*\chi$. Therefore the number operator $n=a^\dagger a=(a^\dagger a+aa^\dagger)/2-1/2$ acquires an additional contribution and this generalizes to all momentum modes.

In the following we will work with a basis of real fields $\chi_1,\chi_2$ defined by 
$\chi(x)=\frac{1}{\sqrt{2}}\big(\chi_1(x)+i\chi_2(x)\big)$, such that $\chi_a(-q)=\chi^*_a(q)$ and 
\begin{equation}\label{4A}
n(\vec{q})+\frac12=\frac{1}{2\Omega_{d+1}}\int_{q_0}\big(\langle\chi^*_1(q)\chi_1(q)\rangle+
\langle\chi^*_2(q)\chi_2(q)\rangle\big).
\end{equation}
The connected part of the two point function describes the propagator ${\cal G}$
\begin{equation}\label{4B}
\langle\chi^*_a(q)\chi_b(q')\rangle={\cal G}_{ab}(q,q')+
\langle\chi^*_a(q)\rangle
\langle\chi_b(q')\rangle.
\end{equation}
For a translation invariant setting ${\cal G}$ is diagonal in momentum space
\begin{equation}\label{4C}
{\cal G}_{ab}(q,q')=\bar G_{ab}(q)\delta(q-q'),
\end{equation}
with $\delta(q-q')=(2\pi)^{d+1}\delta(q_0-q'_0)d^d(\vec q-\vec q\ ')$. Also, translation invariance implies for a possible order parameter $\langle\chi_a(q)\rangle=\sqrt{2}\bar\phi_0\delta(q)\delta_{a1}$ with real $\bar\phi_0$. Here we have chosen the expectation value in the one-direction without loss of generality. We define
\begin{equation}\label{5}
\bar g(q)=\frac12\big(\bar G_{11}(q)+\bar G_{22}(q)\big)
\end{equation}
and express $n(q)$ in terms of the full propagator $\bar G(q)$ and the order parameter $\bar\phi_0$
\begin{eqnarray}\label{6}
n(\vec{q})&=&\int_{q_0}\bar g(q_0,\vec q)+\bar\phi^2_0\delta(\vec{q})-\frac12\nonumber\\
&=&\bar n_p(\vec{q})+\bar\phi^2_0\delta(\vec{q}).
\end{eqnarray}

In the absence of spontaneous symmetry breaking $(\bar\phi_0=0)$ there is no difference between $\chi_1$ and $\chi_2$. Since the expectation values of the type $\langle\chi\chi\rangle$ and $\langle\chi^*\chi^*\rangle$  vanish in this case, it is often convenient to work in the ``complex basis'' of a one component complex field. The propagator is then associated with
\begin{equation}\label{6A}
\langle\chi^*(\vec q\ ')\chi(q)\rangle=\bar G(q)\delta(q-q')
\end{equation}
and $\bar G(q)$ will typically be a complex function. For free bosons one has 
\begin{eqnarray}\label{7}
\bar G_0(q)&=&\left(iq_0+\frac{\vec q\ ^2}{2M}-\sigma \right)^{-1},\nonumber\\
\int_{q_0}\bar G_0(q)&=&T\sum_n
(2\pi in T+\vec q\ ^2/2M-\sigma)^{-1}.
\end{eqnarray}
Performing explicitely the Matsubara sum yields
\begin{equation}\label{7A}
n(\vec{q})=\int_{q_0}\bar G_0(q)-\frac12=\left[\exp\left(\frac{\vec q\ ^2}{2MT}-\frac{\sigma}{T}\right)-1\right]^{-1}
\end{equation}
and we recover the usual occupation number for free bosons. Notice that the subtraction of the term $-1/2$ in eq. (\ref{6}) is crucial for the vanishing of $n(\vec{q})$ for $\vec q\ ^2\to\infty$.

For fermions the situation is similar. There is no fermionic expectation value and the additive constant changes sign according to the anticommuting properties $\frac12(a^\dagger_Fa_F-a_Fa^\dagger_F)=n_F-\frac12$, i.e.
\begin{equation}\label{8}
n_F(\vec{q})=-\int_{q_0}\bar G_F(q_0,\vec q)+\frac12.
\end{equation}
The Matsubara sums for $T\neq 0$ involve now half-integer $n$ and we recover for a free theory the Fermi distribution
\begin{equation}\label{9}
n_F(\vec{q})=\left(\exp\left(\frac{\vec q\ ^2}{2MT}-\frac{\sigma}{T}\right)+1\right)^{-1}.
\end{equation}

In presence of interactions, however, the exact propagator $\bar G(q)$ can be a rather complicated object.   We distinguish between the vacuum and ``occupied states'' where $n(\vec q)$ differs from zero. Often, but not always, an occupied state will go along with a condensate $\bar\phi_0\neq 0$. For non-relativistic interacting particles the propagator becomes complicated only for the occupied states. The complexity of the occupied states is due to the presence of a new scale related to the density. In contrast, the vacuum of nonrelativistic theories is characterized by a simple propagator \cite{Sa}, \cite{CWQP}, even in the  presence of interactions. For the vacuum, we will use a general non-renormalization property of the wave function renormalization in order to clarify some relations between the microscopic formulation of the theory and macroscopic properties. 

Let us consider $T=0$ where the $q_0$ integration is continuous. (More precisely, we define the result of a $q_0$-integral as the principal value and 
$\int_{q_0} f(q_0)=\lim\limits_{\Lambda_0\to\infty}\int^{\Lambda_0}_{-\Lambda_0}\frac{dq_0}{2\pi}f(q_0)$.) 
The vacuum is defined by the limit $n(\vec{q})\to 0$. Since no condensate exists in vacuum $(\bar\phi_0=0)$ we will work in the complex basis. For a stable particle and $\vec q\ ^2>0$ the full vacuum propagator $\bar G(q)$ must have a pole for purely imaginary $q_0$ with positive definite imaginary part. (For free particles the location of the pole is $q_0=i\vec q\ ^2/(2M)$ for $\sigma=0.)$ Often we can write the $q_0$-integration as
\begin{equation}
\int_{q_0}=\frac{1}{2\pi}\int\limits^\infty_{-\infty}dq_0=\frac{1}{4\pi}(\oint_{\rm upper}+\oint_{\rm lower})
\end{equation}
where $\oint_{\rm upper}$ denotes the integral closed on the upper half plane in complex $q_0$-space. As long as the pole remains in the upper half plane the result of the $q_0$ integration is given by the residuum of the pole
\begin{equation}\label{10}
\int_{q_0}\bar G(q_0,\vec q)=\frac12 Z^{-1}_\omega(\vec{q}).
\end{equation}
Here we define the wave function renormalization $Z_\omega(\vec{q})$ by the derivative of the inverse propagator at the location of its zero 
\begin{eqnarray}\label{11}
Z_\omega(\vec{q})&=&-\frac{\partial}{\partial\omega}\bar G^{-1}(\omega,\vec q)_{|\omega_0(\vec{q})},\nonumber\\
\bar{G}^{-1}\big(\omega_0(\vec{q}),\vec q\big)&=&0~,~\omega=-iq_0.
\end{eqnarray}
For a pole in the lower half plane the r.h.s. of eq. (\ref{10}) picks up a minus sign.

Eq. (\ref{10}) constitutes an important relation between the occupation number and the behavior of the full propagator at its pole
\begin{equation}\label{15}
n(\vec{q})=\frac12\big(Z^{-1}_\omega(\vec{q})-1\big).
\end{equation}
From eq. (\ref{15}) we can derive immediately a crucial non-renormalization property for the vacuum of a theory for nonrelativistic particles. Since for the vacuum $n(\vec{q})$ must vanish one concludes that $Z_\omega(\vec{q})$ is not renormalized. It keeps its classical value even in presence of interactions
\begin{equation}\label{16}
\lim\limits_{T\to0,n\to 0}Z_\omega(\vec{q})=1.
\end{equation}

This non-renormalization property resolves several puzzles that would arise if the interaction effects would lead to a nontrivial renormalization $Z_\omega\neq 1$. The first concerns the normalization of the particle number. Indeed, the global $U(1)$-symmetry implies a conserved total particle number $N$, but it does not fix the normalization of $N$. Obviously, $ZN$ is also conserved for arbitrary constant $Z$. The question of the proper normalization of $N$ can be addressed from a microscopic and a macroscopic point of view. In the operator language the normalization of the particle number follows directly form the normalization of the creation and annihilation operators via the commutation relation $[a,a^\dagger]=1$. Constructing the functional integral results in a normalization of the particle number through the normalization of the term $\chi^*\partial_\tau\chi$ in the action. (For an arbitrary normalization of the fields the term linear in the frequency will take the form $iZ_\Lambda(\vec{q}) q_0\chi^*(q)\chi(q)$ instead of the normalization $Z_\Lambda(\vec{q})=1$ employed in eq. (\ref{1}). The particle number reflects then this normalization and the r.h.s of eq. (\ref{4}) is multiplied by a factor $Z_\Lambda(\vec{q})$.) This microscopic point of view relates the particle number to the full propagator $\bar{G}(q)$ for the bare or microscopic fields.

On the other hand, from a macroscopic point of view it should be possible to express all physical quantities relating to macroscopic scales in terms of the $n$-point functions for renormalized fields, without explicit reference to the microscopic formulation. This is a basic concept in quantum field theory and leads to the concept of renormalizable theories when the characteristic macroscopic length scale is many orders of magnitude larger than the microscopic length $\Lambda^{-1}$. In relativistic quantum field theories the renormalized fields $\chi_R=Z^{1/2}\chi$ often involve a wave function renormalization $Z$ than diverges for $\Lambda\to\infty$. Our definition of the particle number seems at first sight to be at odd with this general property of quantum field theory. It is expressed in terms of the bare propagator $\bar{G}(q)$ and not in terms of the two point function for the renormalized field which is $Z\bar{G}(q)$. Therefore the particle number seems to keep some memory of the microscopic normalization. 

In view of eq. (\ref{16}) the resolution of the problem is simple: in vacuum the renormalized fields $\chi_R(q)=Z_\omega(\vec{q})^{1/2}\chi(q)$ are identical to the microscopic fields. Only for nonzero $T$ and $n$ one will find $Z_\omega(\vec{q})\neq 1$. However, this quantity can now be expressed in terms of macroscopic quantities which describe the difference between a thermodynamic equilibrium state and the vacuum. As it should be, no microscopic physics is involved.

A second puzzle is related to a possible dependence of the microscopic wave function renormalization $Z_\Lambda(\vec{q})$ on the scale $\Lambda$ (in some given arbitrary normalization of the field $\chi$). Indeed, the definition of the functional integral does not only involve the action $S$ but also a regularization scale $\Lambda$. Changing $\Lambda$ results in a change of $S$ such that for a given model the action $S_\Lambda$ depends on the choice of $\Lambda$. If for a given normalization of the fields the wave function $Z_\Lambda(\vec{q})$ would run with $\Lambda$, it would not be clear which $Z_\Lambda(\vec{q})$ should be used for the definition of $n$. Fortunately, the non-renormalization property in vacuum (\ref{16}) implies that $Z_{\omega,\Lambda}(\vec{q})= 1$ does not depend on $\Lambda$. 

We close this section by a formal side remark concerning the $q_0$-integral of the vacuum propagator. For large real $q_0$ the renormalization effects vanish such that 
\begin{equation}\label{12}
\lim\limits_{|q_0|\to\infty}\bar{G}^{-1}(q_0,\vec q)=iq_0.
\end{equation}
Using the identity (for $\epsilon>0$)
\begin{equation}\label{13}
\frac12=\int_{q_0}(iq_0+\epsilon)^{-1}
\end{equation}
we can write the $q_0$-integral in a manifestly convergent form
\begin{equation}\label{14}
\int_{q_0}\bar{G}(q_0,\vec q)-\frac12=\int_{q_0}\big\{\bar{G}(q_0,\vec q)-
(iq_0+\epsilon)^{-1}\big\}.
\end{equation}.

\section{Occupation numbers at zero temperature}
\label{low}
For $T=0$ and $n(\vec{q})>0$ we may employ eq. (\ref{15}) (or generalizations thereof) in order to relate the occupation number to the wave function renormalization $Z_\omega(\vec{q})$. Before exploiting this relation in sects. \ref{quadratic} and \ref{density} for bosons, we first discuss the analogous case of fermions. In the absence of spontaneous symmetry breaking one has a single complex propagator $\bar{G}_F$. For a pole in the upper half plane the $q_0$-integration yields
\begin{equation}\label{16B}
n_F(\vec{q})=\frac12\big(1-Z^{-1}_\omega(\vec{q})\big)~,~
Z_\omega(\vec{q})=\frac{1}{1-2n_F(\vec{q})},
\end{equation}
whereas for a pole in the lower half plane one finds 
\begin{eqnarray}\label{16C}
n_F(\vec{q})=\frac12\big(1+Z^{-1}_\omega(\vec{q})\big),~
Z_\omega(\vec{q})=\frac{1}{2n_F(\vec{q})-1}.
\end{eqnarray}
For a free theory and $\sigma>0$ the pole remains in the upper half plane for $\vec q\ ^2>k^2_F~,~k^2_F=2M\sigma$. For these  momenta $n(\vec{q})=0$ implies $Z_\omega(\vec{q})=1$. In contrast, for $\vec q\ ^2<k^2_F$ the pole is in the lower half plane. Now $n(\vec{q})=1$ results again in $Z_\omega(\vec{q})=1$. Thus a sharp Fermi surface is equivalent to $Z_\omega(\vec{q})=1$. In contrast, the presence of interactions can ``smoothen'' the Fermi surface and lead to $Z_\omega(\vec{q})\neq 1$, even for $T=0$. 

The jump in the particle number $n_F(\vec{q})$ is an interesting example how quantum mechanical discreteness arises from a perfectly continuous functional integral. Let us for simplicity consider a finite box such that the momenta are discrete. (This discreteness in the number of integration variables is not related to quantum mechanics - it is the same for classical physics.) The integration measure in the functional integral (\ref{1}) depends continuously on $\sigma$. One may therefore expect that also the expectation values like the average value $n_F(\vec{q})$ depend continuously on $\sigma$. We have just demonstrated that this is not the case for $T=0$. If for a free theory $k^2_F$ crosses the value $\vec q\ ^2$ for one of the discrete momentum levels the pole in the propagator moves from the upper half plane to the lower half plane and $n_F(\vec{q})$ jumps by one unit. For stable particles this generalizes to the interacting theory. The number of particles with a given momentum $\vec q$ is discrete!

For bosons the situation at $T=0, ~n\neq 0$ is more subtle due to the presence of a Bose-Einstein condensate, $\bar\phi_0\neq 0$. We now employ the ``real basis'' of fields $\chi_1$ and $\chi_2$. In principle, one has to consider separately the poles of the propagator $\bar{G}_{11}$ for the radial mode and of $\bar{G}_{22}$ for the Goldstone mode. Let us denote by $\bar G$ the $2\times 2$ matrix with elements $\bar{G}_{ab}$ and investigate the inverse matrix $\bar P=\bar G^{-1}$. By Euclidean time reversal symmetry it has the general form
\begin{equation}\label{16D}
\bar P=\left(\begin{array}{cc}
a,&-q_0Z_\phi\\q_0Z_\phi,&b\end{array}\right)
\end{equation}
with $a,b,Z_\phi$ depending only on $\bar q^2$ and $q^2_0$. This implies that $\bar{g}$ is an even function of $q_0$
\begin{equation}\label{16E}
\bar G=(ab+q^2_0Z_\phi^2)^{-1}
\left(\begin{array}{cc}
b,&q_0Z_\phi\\-q_0Z_\phi,&a\end{array}\right)~,~
\bar g=\frac12\frac{a+b}{ab+q^2_0Z_\phi^2}.
\end{equation}
Therefore the location of the poles for $\bar{G}_{11}$ and $\bar{G}_{22}$ is the same, namely for 
\begin{equation}\label{16F}
q_0=\pm i\sqrt{\frac{ab}{Z_\phi^2}}~,~\omega_0=\pm \sqrt{\frac{ab}{Z_\phi^2}}.
\end{equation}
Here we use that $a,b,Z_\phi^2$ are real and positive and we assume $Z_\phi>0$. We recall that $a,b,Z_\phi$ may depend on $q^2_0$ such eq. (\ref{16F}) is an implicit equation for $q_0(\vec q)$. 

We notice that in the real basis $\bar{g}$ has poles both in the upper and lower half plane. There residua have an opposite sign such that both contribute equally to $\int_{q_0}\bar{g}(q)$. If the solution of eq. \eqref{16F} is unique we define
\begin{equation}\label{16G}
Z_\omega(\vec{q})=-\frac12\frac{\partial \bar{g}^{-1}(\omega,\vec q)}{\partial\omega}_{|\omega_0(\vec{q})>0}=
\frac12\frac{\partial\bar{g}^{-1}(\omega,\vec q)}{\partial\omega}_{|\omega_0(\vec{q})<0}.
\end{equation}
This normalization ensures consistency with the complex basis in the absence of spontaneous symmetry breaking. Relations of the type (\ref{10}) and (\ref{15}) continue to hold for $\omega_0>0$. The residuum of the pole at $\omega_0>0$ is given by $Z^{-1}_\omega$, whereas the one for the pole at $\omega_0<0$ has the opposite sign. In consequence, both poles contribute an equal amount to the $q_0$-integral.  This yields for bosons at $T=0$ and $\vec q\ ^2>0$
\begin{equation}\label{16H}
Z_\omega(\vec{q})=\frac{1}{1+2n(\vec{q})}.
\end{equation}
Thus for single poles we recover eq. \eqref{15}, while generalizations occur if eq. \eqref{16F} has more than one solution. 

\section{Boson occupation numbers for linear dynamic term}
\label{Bosonoccupationnumbers}
In this section, we compute the occupation numbers for bosons at $T=0$ in the approximation that the $q_0$-dependence of the inverse propagator is linear, as for the classical action. This amounts to an ansatz where the coefficients $a,b,Z_\phi$ in eq. (\ref{16D}) depend on $\vec q$, but not on $q_0$. For $Z_\phi=1,~a=\vec q\ ^2/(2M)+2\lambda_\phi\bar\phi^2_0,~b=\vec q\ ^2/(2M)$ this amounts to the Bogoliubov approximation. 
We will encounter infrared problems in low dimensions $(d=1)$. We show in the next section how those are cured by a more general form of the propagator. In fact, for $d=1,2$ the $q_0$-dependence of the inverse propagator becomes quadratic in the regime with nonzero density, where the fluctuations of the Goldstone bosons dominate \cite{CWQP}. We also encounter ultraviolet problems if $Z_\phi$ differs from one. They are cured by a renormalization group improved treatment that we discuss in sects. \ref{improvement} to \ref{density}. These issues show that a functional integral computation of occupation numbers is not trivial and needs a thorough understanding of the many body dynamics.

In the limit of $q_0$-independent $a,b$ and $Z_\phi$ one finds
\begin{equation}\label{16I}
Z_\omega=\frac{2Z^2_\phi|\omega_0|}{a+b}=\frac{2Z_\phi\sqrt{ab}}{a+b}.
\end{equation}
In the absence of spontaneous symmetry breaking one has $a=b$ and therefore $Z_\omega=Z_\phi$ (we assume $Z_\phi>0$). In this case possible deviations of $Z_\omega$ from one can only arise from $Z_\phi\neq1$. In the presence of spontaneous symmetry breaking, however, $Z_\omega$ deviates from one even for $q_0$-independent $a,b$  and for $Z_\phi=1$. We may parameterize
\begin{equation}\label{16J}
a=\bar A\left(\frac{\vec q\ ^2}{2M}+2\lambda_\phi \bar\phi_0^2\right)~,~b=\bar A\frac{\vec q\ ^2}{2M}
\end{equation}
where $\bar A$ and $\lambda_\phi$ may depend on $\vec q\ ^2$. This reflects the vanishing of $a-b$ for $\bar\phi_0\to 0$ and the Goldstone-character of $\chi_2$, i.e. $\lim\limits_{\vec q\ ^2\to 0}b=0$. We will employ later that $\bar P$ is related to the second functional derivative of the effective action $\Gamma$. The coefficients (\ref{16J}) follow from a simple form of $\Gamma$, i.e.
\begin{equation}\label{33A}
\Gamma=\int_x \left\{Z_\phi\bar\phi^*\partial_\tau\bar\phi-\bar A\bar\phi^*\frac{\Delta}{2M}\bar\phi
+\frac{\bar A}{2}\lambda_\phi(\bar\phi^*\bar\phi-\bar\phi^2_0)^2\right\}.
\end{equation}
They can also arise from a more general form of $\Gamma$. 

With the ansatz (\ref{16D}) (\ref{16J}) we find for  $\vec q\ ^2>0$
\begin{eqnarray}\label{16K}
n_p(\vec{q})&=&\frac12(Z_\omega^{-1}-1)=\frac12
\left\{\frac{1}{|Z_\phi|}\sqrt{1+\frac{(a-b)^2}{4ab}}-1\right\}\nonumber\\
&=&\frac12\Bigg\{\frac{1}{Z_\phi}\Bigg(1+
\frac{\lambda^2_\phi\bar\phi_0^4}{\frac{\vec q\ ^2}{2M}\left(\frac{\vec q\ ^2}{2M}+2\lambda_\phi\bar\phi_0^2\right)}\Bigg)^{1/2}-1.\Bigg\}
\nonumber\\
\end{eqnarray}
For $Z_\phi=1$ and constant $\lambda_\phi$ this is the occupation number for the Bogoliubov theory. Writing the density $n$ as a condensate density $\bar\phi^2_0$ and a ``particle density''  $\bar n_p$
\begin{equation}\label{34A}
n=\bar\phi^2_0+\bar n_p~,~\bar n_p=\int_{\vec q}n_p(\vec q)
\end{equation}
the Bogoliubov theory yields a finite result for $\bar n_p$ only for $d>1$. Indeed, for small $q$ the occupation number diverges $\sim q^{-1}$ in the Bogoliubov theory
\be\label{37A}
n_p(\vec q)=\frac{\sqrt{M\lambda_\phi}\bar\phi_0}{2q}.
\ee
For $d=2,3$ the particle density \eqref{34A} is dominated by modes with momenta $\vec q^2\approx 4M\lambda_\phi\bar\phi^2_0$. 

We will see that quantum fluctuations are responsible for important modifications of $n_p(\vec q)$ for small $q=|\vec q|$. The dominant effect concerns the $q_0$ dependence of the propagator discussed in the next section. In this section we first retain the linear dependence of $\bar G^{-1}$ on $q_0$ and show that the dependence of $\lambda_\phi$ and $Z_\phi$ on $\vec q$ does not resolve the infrared problems. Evaluating eq. \eqref{16K} for $\vec q\ ^2\to 0$ the occupation number diverges according to
\begin{equation}\label{16L}
n_p(\vec{q})=\frac{\sqrt{M\lambda_\phi(\vec{q})}}{2Z_\phi(\vec{q})}\frac{\bar\phi_0}{|\vec q|}.
\end{equation}
The detailed behavior depends on the $\vec q$-dependence of $\lambda_\phi$ and $Z_\phi$. This can be extracted from a recent  functional renormalization group study of quantum phase transitions \cite{CWQP}. For the far infrared behavior $\vec q^2\to 0$ the Goldstone regime applies. The relevant physics strongly depends on the space dimension $d$. For $d=3$ one finds that $Z_\phi$ vanishes logarithmically with $q=|\vec q|$
\begin{equation}\label{N1}
Z_\phi\sim \ln^{-1}
\left(\frac{k^2_0}{q^2}\right).
\end{equation}
(In the notation of \cite{CWQP} one has $Z_\phi=\bar S=AS$ if appropriate units for length and time are used.) Also $\lambda_\phi$ vanishes logarithmically, $\lambda_\phi\sim Z_\phi$. The overall behavior for $d=3$
\begin{equation}\label{N2}
n(\vec q)\sim
\frac{\ln^{\frac12}(k^2_0/q^2)}{q}
\end{equation}
differs from the free Bose-Einstein distribution which vanishes for $q\neq 0$ and $T=0$. The infrared contribution to the total particle number $\sim\int dqq^2n(q)$ remains finite.

For $d=1$ and $d=2$ the issue is more involved. It is convenient to introduce renormalization fields
\be\label{40Aa}
\phi=\sqrt{\bar A}\bar\phi.
\ee
The renormalized order parameter $\phi_0$ is related to the superfluid density $n_s$ by $n_s=\rho_0=\phi^2_0$. For $T=0$ the Galilei invariance for non-relativistic bosons yields a Ward identity which implies that the total particle density $n$ equals $\rho_0$. Since $\bar A$ depends on momentum the unrenormalized order parameter $\bar\phi_0$ becomes effectively momentum dependent. In particular, $\bar\phi_0$ vanishes for $d=1$ and $q\to 0$, while $\phi_0$ remains different from zero. Writing
\begin{equation}\label{N3}
Z_\phi=\bar A S~,~S\sim q^{-\eta_S}~,~\bar A\sim q^{-\eta}
\end{equation}
one finds $\eta_S\leq -1$ for $d=1,2$. For $d=2$ the anomalous dimension $\eta$ vanishes for $q\to 0$, while for $d=1$  one has a range $0<\eta\leq 1/2$. We also introduce the renormalized coupling $\lambda=\lambda_\phi/\bar A$ and note that $\sigma_s=S/\lambda$ goes to a constant for $q\to 0$. 

The resulting expression for the occupation numbers reads for $q\to 0$
\be\label{40B}
n(\vec q)=\left(\frac{M\rho_0}{\sigma_s}\right)^{1/2}\frac{1}{2\sqrt{S(q)}\bar A(q)q}\sim
q^{-1+\eta+\frac{\eta_S}{2}}
\ee
For $d=2$ the integral $\int dq q^{d-1}n(q)$ remains infrared finite only for $\eta_S>-2$, while for $d=1$ finiteness requires $\eta_S>-2\eta$. For $d=1$ this condition is not obeyed - in this case already the ``classical approximation'' with vanishing $\eta$ and $\eta_S$ leads to a logarithmic divergence. We will see in the next section how this too strong infrared increase of $n(\vec q)$ is cured by a modification of the $q_0$-dependence of the propagator.

We next turn to the behavior for large $q$ as given (for all dimensions) by eq. (\ref{16K}). We may use $Z_\phi=1+\Delta Z_\phi$ where $\Delta Z_\phi$ vanishes for $q\to \infty$. This results in 
\begin{equation}\label{N11}
n(\vec q)=
\frac{M^2\lambda^2_\phi\bar\phi^4_0}{q^2(q^2+4M\lambda_\phi\bar\phi^2_0)}-\frac{\Delta Z_\phi}{2}.
\end{equation}
For a first evaluation of $\Delta Z_\phi$ we may use the flow equation of \cite{CWQP}, replacing $k$ by $\gamma q$. For the regime $q^2\gg 4M\lambda_\phi\bar\phi^2_0$ one finds
\begin{equation}\label{N12}
q\frac{\partial}{\partial q}Z_\phi=-(\eta+\eta_S)Z_\phi
=-\frac{32v_d}{d}M^2\lambda^2_\phi\bar\phi^2_0(\gamma q)^{d-4}
\end{equation}
with $v^{-1}_d=2^{d+1}\pi^{d/2}\Gamma(d/2)$. In consequence, the leading order (with constant $\lambda_\phi,\bar\phi_0$)
\begin{equation}\label{N13}
\Delta Z_\phi=\frac{32 v_d M^2\lambda^2_\phi\bar\phi^2_0}{d(4-d)}(\gamma q)^{d-4}
\end{equation}
is positive and decreases $\sim q^{d-4}$. A similar result follows from a direct evaluation of the one loop contribution to $Z_\phi$, if $Z_\phi$ is defined as the coefficient of the term linear in $q_0$ of an expansion of the inverse propagator in powers of $q_0$ - cf. appendix A.

The simple estimate \eqref{N11} for $\Delta Z_\phi$ cannot be true. The decay for large $q$ is too slow. The term $\sim\Delta Z_\phi$ would dominate $n(\vec q)$ for large $q$ and lead to negative occupation numbers. Also the $\vec q$-integral for the total particle density would not converge for $d>2$. The origin of this problem is again the neglection of the non-linear $q_0$ dependence in the inverse propagator. It is not justified to evaluate the $q_0$-dependence by the $q_0$-derivative of the inverse propagator at $q_0=0$. If one wants to use an approximation linear in $q_0$ for the inverse propagator, $Z_\phi$ should be evaluated at the location of the pole rather than near $q_0=0$. 
We will develop in sects. \ref{improvement} to \ref{density} a renormalization group treatment for the computation of $n(\vec q)$. As a result, one finds that there is actually no contribution from $\Delta Z_\phi$ in eq. \eqref{N11}. Omitting this term the ultraviolet contribution to the total particle number becomes UV-finite
\be\label{41A}
n_{UV}=\int_{\vec q}
\frac{M^2\lambda^2_\phi\bar\phi_0^4}{q^2(q^2+4M\lambda_\phi\bar\phi^2_0)},
\ee
as for the Bogoliubov approximation. 

\section{Boson occupation numbers for general dynamic term}
\label{dynamicterm}
Let us next address the infrared problem for $d=1$. For the Bogoliubov approximation the occupation numbers diverge $\sim 1/q$ and the particle density $n$ does not remain finite. This problem is exacerbated if the $\vec q$-dependence of $\lambda_{\phi}\bar \phi_0^2/Z^2_\phi$ is taken into account. The remedy of this  problem is related to the $q_0$-dependence of $a$ and $b$ in eq. (\ref{16D}), that was neglected so far. Indeed, the  investigation of \cite{CWQP} reveals a crossover to a ``relativistic kinetic term'' for small  momenta. For $q\ll q_r$ one may neglect the off-diagonal term $\sim q_0 Z_\phi$ in eq. (\ref{16D}).  
Due to the strong decrease of $Z_\phi$ for small $q$ (\ref{N3}) the $q_0$-dependence is now dominated by a contribution quadratic in $q^2_0$ in $a$ and $b$
\begin{equation}\label{N4}
a=\bar A\left(\frac{\vec q\ ^2}{2M}+2\lambda_\phi\bar\phi^2_0\right)
+\bar V q^2_0~,~b=\frac{\bar A\vec q\ ^2}{2M}+\bar V q^2_0.
\end{equation}
In this region the Goldstone mode dominates
\begin{equation}\label{N5}
\bar g\approx\frac12\bar G_{22}\approx\frac{1}{2b}
\end{equation}
and the location of the pole is now at
\begin{equation}\label{N6}
\omega_0=\pm \sqrt{\frac{\bar A q^2}{2M\bar V}}.
\end{equation}
With 
\begin{equation}\label{N7}
Z_\omega=2\bar V|\omega_0|=
\left(\frac{2\bar A\bar V}{M}\right)^{1/2}q
\end{equation}
the particle density for $q\to 0~(q\neq 0)$ obeys 
\begin{equation}\label{N8}
n(\vec q)=\left(\frac{M}{8\bar A\bar V}\right)^{1/2}q^{-1}.
\end{equation}

The result \eqref{N8} for the low momentum behavior of the occupation numbers differs qualitatively from the Bogoliubov theory \eqref{37A}. It involves the new coupling $\bar V$, such that the coefficient in front of $q^{-1}$ does not vanish anymore for $\bar\phi_0\to 0$. For $q\to 0$ one finds \cite{CWQP} that $V=\bar V/\bar A$ goes to a constant such that (cf. eq. (\ref{N3})) for $d=1,2$
\begin{equation}\label{N9}
n(\vec q)\sim q^{-(1-\eta)}.
\end{equation}
In consequence, the integral $\int dq q^{d-1}n(q)$ remains infrared convergent both for $d=2$, where $\eta=0$, and for $d=1$, where $\eta>0$. We note that the infrared convergence for $d=1$ arises only as a consequence of a nonvanishing anomalous dimension $\eta$. The same effect guarantees that the correlation function in position space decays with a power law for large separation $(d=1,\tau=it)$
\begin{equation}\label{N10}
\bar G\sim (\vec r\ ^2+v^2\tau^2)^{-\frac\eta2}.
\end{equation}

Away from the infrared limit $q\to 0$ one has to follow the location of the poles of the general propagator given by eqs. (\ref{N4}), (\ref{16E}). They obey, with $\lambda=\lambda_\phi/\bar A~,~ \rho_0=\bar A\bar\phi^2_0~,~\lambda\rho_0=\lambda_\phi\bar \phi^2_0,~S=Z_\phi/\bar A$,
\ba\label{48A}
q^2_0&=&-\frac{1}{2V^2}\left[S^2+2V\lambda\rho_0+V\frac{\vec q^2}{M}\right.\nn\\
&&\pm\left\{(S^2+2V\lambda\rho_0)^2+2S^2V\frac{\vec q^2}{M}\right\}^{1/2}\Big].
\ea
We discuss here the case where $S,V$ and $\lambda$ are independent of $q_0$. (A generalization for mildly $q_0$-dependent couplings will result in implicit equations where these quantities have to be evaluated on the location of the poles.) On the other hand, the couplings depend on $q$. For large $q$ the fluctuation effects are small, such that $V\to 0,S\to 1$, while for small $q$ one has $S\to 0, V\neq 0$. All solutions of eq. (\ref{48A}) correspond to negative $q^2_0$ and therefore to pairs of purely imaginary $q_0$. 

Let us start with $V=0$ where the poles at
\be\label{48B}
q_0=\pm iS^{-1}\sqrt{\frac{\vec q^2}{2M}\left(\frac{\vec q^2}{2M}+2\lambda\rho_0\right)}
\ee
can be associated with the Goldstone mode. We observe positive and negative frequencies for the excitations, corresponding to particles and antiparticles, the latter being associated to ``holes'' in the condensate. For small nonzero $V$, the poles (\ref{48B}) get slightly shifted. Two new poles appear at large $|q_0|$, 
\be\label{48C}
q_0=\pm i\frac{S}{V}.
\ee
As $S/V$ decreases towards zero, these modes correspond to the radial modes of the relativistic model, i.e. for $S\to 0$ 
\be\label{48D}
q^2_{0,R}=-\frac{1}{V}\left[\frac{\vec q^2}{2M}+2\lambda\rho_0+\frac{S^2}{V}\left(1+\frac{\vec q^2}{4M\lambda\rho_0}\right)\right].
\ee
The Goldstone poles take for $S\to 0$ the standard relativistic values 
\be\label{48E}
q^2_{0,G}=-\frac{\vec q^2}{2MV}\left(1-\frac{S^2}{2V\lambda\rho_0}\right).
\ee
\\
The radial mode corresponds to collective fluctuations related to the size of the condensate $\rho_0$, whereas the Goldstone mode is associated to phase changes of the condensate. More generally, we refer to the negative (positive) sign of the square root in eq. (\ref{48A}) as the Goldstone (radial) poles. 

For a computation of the occupation number with the propagator \eqref{N4} we evaluate for $\vec q\neq 0$
\ba\label{59A}
\hat n(\vec q)&=&\int_{q_0}\bar g(q_0,\vec q)=\frac12\int_{q_0}tr\bar G(q_0,\vec q)\nn\\
&=&\frac12\int_{q_0}\frac{a+b}{ab+Z^2_\phi q^2_0}\nn\\
&=&\frac{1}{\bar A}\int_{q_0}
\left(\frac{q^2}{2M}+Vq^2_0+\lambda\rho_0\right)\det\nolimits^{-1}(q)
\ea
with 
\ba\label{59B}
&&\det(q)=\left(\frac{q^2}{2M}+Vq^2_0+2\lambda\rho_0\right)
\left(\frac{q^2}{2M}+Vq^2_0\right)+S^2q^2_0\nn\\
&&=V^2\left(q^2_0+\frac{1}{2V^2}(\alpha+\sqrt{\beta})\right)
\left(q^2_0+\frac{1}{2V^2}(\alpha-\sqrt{\beta})\right),\nn\\
\ea
and
\ba\label{59C}
\alpha&=&S^2+2V\lambda\rho_0+\frac{Vq^2}{M},\nn\\
\beta&=&(S^2+2V\lambda\rho_0)^2+2S^2V\frac{q^2}{M}.
\ea
Writing
\be\label{59D}
\det\nolimits^{-1}(q)=\frac{1}{\sqrt{\beta}}
\left(\frac{1}{q^2_0+\frac{\alpha-\sqrt{\beta}}{2V^2}}-
\frac{1}{q^2_0+\frac{\alpha+\sqrt{\beta}}{2V^2}}\right).
\ee
we recognize the contribution from the radial and Goldstone poles
\ba\label{59E}
\hat n(\vec q)=\frac{1}{2\bar A V}\int_{q_0}
\left(\frac{1-S^2/\sqrt{\beta}}{q^2_0+\frac{\alpha-\sqrt{\beta}}{2V^2}}+
\frac{1+S^2/\sqrt{\beta}}{q^2_0+\frac{\alpha+\sqrt{\beta}}{2V^2}}\right).
\ea
The $q_0$-integration is performed easily
\ba\label{59F}
\hat n (\vec q)&=&\frac{1}{2\sqrt{2}A}
\left[\frac{1-S^2/\sqrt{\beta}}{\sqrt{\alpha-\sqrt{\beta}}}
+\frac{1+S^2/\sqrt{\beta}}{\sqrt{\alpha+\sqrt{\beta}}}\right]\nn\\
&=&\frac{M}{2\sqrt{2}\bar AV}
\big\{q^2(q^2+4M\lambda\rho_0)\big\}^{-1/2}{\cal F} ,
\ea
with
\ba\label{59G}
{\cal F} &=&\sqrt{\alpha-\sqrt{\beta}}+\sqrt{\alpha+\sqrt{\beta}}\nn\\
&&+\frac{S^2}{\sqrt{\beta}}
\left(\sqrt{\alpha-\sqrt{\beta}}-\sqrt{\alpha+\sqrt{\beta}}
\right).
\ea
For the limiting case $V\to 0$ one has 
$\alpha=\sqrt{\beta}\to S^2+V(q^2/M+2\lambda\rho_0),{\cal F} \to \sqrt{2}(V/S)(q^2/M+2\lambda\rho_0)$ and we recover eq. \eqref{16K}. In the limit $S\to 0$ we can use $\alpha=(q^2+2M\lambda\rho_0)V/M,\sqrt{\beta}=2V\lambda\rho_0$ and ${\cal F} =\sqrt{V/M}(\sqrt{q^2}+\sqrt{q^2+4M\lambda\rho_0})$ such that we recover \eqref{N8}.

For an estimate of the leading behavior for $q\to \infty$ we need ${\cal F} /V$ and therefore the $q$-dependence of $V$. A renormalization group estimate \cite{CWQP} yields
\be\label{59H}
\lim_{q\to \infty}V(q)=\frac{10 v_d}{d(6-d)}\lambda^2\rho_0q^{d-6}.
\ee
In consequence, the terms $\sim Vq^2$ in eq. \eqref{59C} become subleading and the ultraviolet behavior is well described by the linear dynamic term in sect. \ref{Bosonoccupationnumbers}. 

In the spirit of a renormalization group improvement, which will be motivated in the next sections, we may derive a differential equation for the $q$-dependence of $n_p$
\ba\label{68X}
\frac{\partial n_p(q)}{\partial \ln q}&=&
\frac{\partial\hat n(q)}{\partial \ln q}=\gamma(q),\nn\\
\gamma(q)&=&-\frac{M}{2\sqrt{2}\bar A V}
\big\{q^2(q^2+4M\lambda\rho_0)\big\}^{-1/2}\nn\\
&&\left\{\left(1+\frac{q^2}{q^2+4M\lambda\rho_0}\right){\cal F}-\frac{\partial{\cal F}}{\partial\ln q}\right\}.
\ea
Here $\partial{\cal F}/\partial\ln q$ should be evaluated for fixed couplings. The advantage of such an equation is that the ultraviolet problem associated with the additive counterterm is now absent. For large $q$ the function $\gamma(q)$ decays $\sim q^{-4}$. We can therefore integrate eq. \eqref{68X}, starting with an initial value for large $q=\Lambda$ for which the perturbative result eq. \eqref{16K} with $Z_\phi=1$) can be trusted. Furthermore, the ``renormalization group improvement'' of using $q$-dependent couplings $\bar A,S,\lambda_\phi,\bar\phi_0,V$ will make the solution of eq. \eqref{68X} rather accurate. It can be used for all dimensions and is free of infrared or ultraviolet problems.

Eq. \eqref{68X} can directly be used to compute the particle density or condensate depletion 
$n_p=\int_{\vec q}n_p(\vec q)$ by solving a second order differential equation. With $x=\ln (q/\hat k)$ and $\hat k$ a suitable momentum unit, one may compute $n_p=\hat k^d\tilde n(x\to-\infty$) from the differential equation
\be\label{68Y}
\frac{\partial^2\tilde n}{\partial x^2}-d
\frac{\partial\tilde n}{\partial x}=v_d e^{dx}\gamma(x),
\ee
using the initial condition $\tilde n(\infty)=0$ (or a similar condition for $x=\ln(\Lambda/\hat k)$).

\section{Renormalization group improvement}
\label{improvement}
In view of the rather complicated momentum dependence of the full propagator it is useful to consider the contributions from fluctuations with different momenta $\vec p$ separately. This will also allow us to use a renormalization group improvement by employing effective couplings appropriate for a given momentum range. In this section we will follow a heuristic approach. The second part of this paper will present a more formal development, based on exact functional renormalization group equations.

We start by adding to the inverse propagator an infrared cutoff piece such that
\begin{equation}\label{B1}
\bar G_k(q_0,\vec q)=\big(\bar P(q_0,\vec q)+R_k(\vec q)\big)^{-1}.
\end{equation}
The infrared cutoff is a $2\times 2$ diagonal matrix and suppresses all modes with $\vec q\ ^2<k^2$. We choose 
\begin{equation}\label{B2}
R_k(\vec q)=\frac{\bar A}{2M}(k^2-\vec q\ ^2)\theta (k^2-\vec q\ ^2)
\end{equation}
such that the modes with $\vec q\ ^2>k^2$ are not affected. On the other hand, the increase of the classical free propagator for small momenta, $G\sim q^{-2}$, is stopped at $q^{-2}=k^{-2}$. We define a $k$-dependent occupation number (for $\vec q\neq 0)$ 
\begin{equation}\label{B3}
n_{p,k}(\vec q)=\frac12\int_{q_0}Tr\bar G_k(q)-\frac12
\end{equation}
and consider the derivative
\begin{equation}\label{B4}
\partial_k n_{p,k}(\vec q)=-\frac12\int_{q_0}Tr
\big\{\partial_kR_k(\vec q)\bar G^2_k(q_0,\vec q)\big\}.
\end{equation}
We note that $\partial_k R_k$ vanishes for $\vec q\ ^2>k^2$, such that the running stops once $k\leq q$. Furthermore, for the classical propagator the r.h.s. of eq. (\ref{B4}) becomes independent of $\vec q$ for $\vec q\ ^2<k^2$. The $q_0$-integration on the r.h.s. of eq. \eqref{B4} will be dominated by $k$-dependent poles. The physical occupation number obtains by solving eq. (\ref{B4}) for $k\to 0$. Indeed, the absence of the cutoff for $k=0$ implies 
$n_p(\vec q)=n_{p,0}(\vec q)$. 

More explicitely, the flow equation reads
\begin{eqnarray}\label{B5}
&&\partial_k n_{p,k}(\vec q)=-\frac{k\bar A}{2M}\theta (k^2-\vec q\ ^2)\left(1-\frac\eta2\right)\nonumber\\
&&~ \hspace{1.4cm}\int_{q_0}(\bar G^2_{11}+\bar G^2_{22}+2\bar G_{12}\bar G_{21}).
\end{eqnarray}
We will neglect the anomalous dimension
\begin{equation}\label{B6}
\eta=-k\frac{\partial}{\partial k}\ln\bar A
\end{equation}
in this section. Inserting the ansatz (\ref{16D}) and approximating $a,b$ and $Z_\phi$ to be $q_0$-independent one finds
\begin{equation}\label{B7}
k\partial_kn_{p,k}(\vec q)=-
\frac{4M^2\lambda^2_\phi\bar\phi^4_0}{Z_\phi k(k^2+4M\lambda_\phi\bar\phi^2_0)^{3/2}}\theta(k^2-\vec q\ ^2).
\end{equation}
Using for $k\to\infty$ the initial condition $n_{p,k\to\infty}=0$ an integration for $k$-independent $\lambda_\phi,\bar\phi_0,Z_\phi$ yields for $k=0$ eq. (\ref{16K}). 

The renormalization group improvement uses $k$-dependent parameters $\lambda_\phi,\bar\phi_0$ and $Z_\phi$. These $k$-dependent couplings are in turn computed in presence of the cutoff scale $k$, such that they approximate best the relevant couplings in the momentum range $\vec p\ ^2\approx k^2$. We will motivate this procedure more explicitely in the second part of the paper. A crucial property for the ultraviolet issue is $Z_\phi(k\to \infty)=1$. This simply follows from the absence of fluctuation effects for an infinitely large cutoff. This property guarantees the correct initial value for the flow $n_{p,k\to \infty}(q)=0$ for all $\vec q$. Indeed, the classical contribution to $n_{p,k}$ for very large $k$ yields eq. (\ref{16K}) with $\vec q\ ^2$ replaced by $k^2$ and all parameters evaluated at the scale $k$. It is obvious that  $Z_\phi(k\to\infty)=1$ is needed for a vanishing occupation $n_{p,k}$ in this limit. The ``initial value'' $Z_\phi(k\to\infty)=1$ is well compatible with a nonzero value $Z_\phi(k\to 0)$. While only the latter is used in eq. (\ref{16K}), the flow equation \eqref{B4} involves the value of $Z_\phi(k)$ for all $k^2>\vec q\ ^2$. We will see that the approach with a varying $Z_\phi(k)$ avoids in a simple way the difficult ultraviolet issues encountered in the direct evaluation of eq. (\ref{16K}). 

Let us consider the range of high momenta $\vec q\ ^2\gg 4M\lambda_\phi\bar\phi^2_0$. In the relevant $k$-range  $k^2\geq \vec q\ ^2$ we can approximate
\begin{equation}\label{B8}
\partial_kn_p(\vec q)=-4M^2\lambda^2_\phi(k)\bar\phi^4_0(k)Z^{-1}_\phi(k)k^{-5}\theta(k^2-\vec q\ ^2).
\end{equation}
This yields
\begin{equation}\label{B9}
n_p(\vec q)=M^2\langle\lambda^2_\phi\bar\phi^4_0Z^{-1}_\phi\rangle q^{-4}
\end{equation}
where we have defined the weighted average
\begin{equation}\label{B10}
\langle \lambda^2_\phi\bar\phi^{-4}_0Z^{-1}_\phi\rangle=
\int\limits^\infty_qdk\lambda^2_\phi(k)\bar\phi^4_0(k)Z^{-1}_\phi(k)k^{-5}~ {\Big/}~
\int\limits^\infty_qdkk^{-5}.
\end{equation}
The factor $k^{-5}$ strongly enhances the weight of the range $k\approx q$ and we may approximate
\begin{equation}\label{B11}
\langle\lambda^2_\phi\bar\phi^4_0Z^{-1}_\phi\rangle\approx\lambda^2_\phi(q)\bar\phi^4_0(q)Z^{-1}_\phi(q).
\end{equation}
Comparison with eq. (\ref{16K}), (\ref{N11}) yields $\Delta Z_\phi=0$ and therefore effectively $Z_\phi=1$ for the range $q^2\gg 2M\lambda^2_\phi\bar\phi^2_0$. No difficult ultraviolet issue needs to be addressed anymore. In particular, the renormalization group improvement avoids the need to evaluate the propagator at nonzero $q_0$, corresponding to the location of the pole. We have also found that the parameters $\lambda_\phi,\bar\phi^2_0,Z_\phi$ appearing in eq. (\ref{16K}) should be replaced by $q$-dependent effective couplings. We emphasize that our argument in favor of $\Delta Z_\phi(q)=0$ is valid for $q^2\gg 2M\lambda^2_\phi\bar\phi^2_0$ and does not necessarily apply to small $q^2$. 

\section{Occupation numbers from source terms}
\label{sourceterms}
In sects. \ref{flow}-\ref{density} we address the formulation of the flow equations for $k$-dependent occupation numbers more formally. We will derive an exact functional differential equation and propose non-perturbative approximations. The solution of the flow equations offers an alternative to the direct computation of $n(\vec q)$ from the propagator \eqref{6}. The $q_0$-integration in eq. \eqref{6} depends in a critical way on the precise knowledge of the $q_0$-dependence of the propagator which is quite involved. In contrast, the flow equations are less sensitive to this dependence. Only a small $q_0$ range influences the running at a given renormalization scale and rather crude approximations are often sufficient. The prize to pay is the numerical integration of the flow equations. In this paper we will not perform detailed quantitative computations. We rather use the flow equations for a justification of the renormalization group improvement discussed in sect. \ref{improvement}.

In the present short section we set up the formal tools for the computation of the momentum distribution $n(\vec{q})$. For this purpose we add appropriate source terms to the action (\ref{1})
\begin{equation}\label{17}
S\to S-\int_qh(\vec{q})\chi^*(q)\chi(q).
\end{equation}
The partition function $Z$ depends then on $h(\vec{q})$ and we can formally extract $n(\vec{q})$ as
\begin{equation}\label{18}
n(\vec{q})\pm\frac12=\Omega^{-1}_{d+1}\frac{\delta}{\delta h(\vec{q})}\ln Z_{|h=0}.
\end{equation}
Here we set the source term to zero after the differentiation and the minus sign applies for fermions. We note that the source multiplies a composite operator and is nonlocal in the fields $\chi(x)$
\begin{eqnarray}\label{19}
&&\int_q h(\vec{q})\chi^*(q)\chi(q)\nonumber\\
=&&\int_\tau\int_{\vec x}\int_{\vec y}
\tilde h(\vec x-\vec y)\chi^*(\tau,\vec x)\chi(\tau,\vec y),
\end{eqnarray}
with $\tilde h(\vec z)$ the Fourier transform of $h(\vec{q})$. Eq. (\ref{19}) shows that the source term remains translation invariant. 

We will conveniently work with the effective action $\Gamma[\bar{\phi}]$ which generates the $1 PI$ correlation functions. It obtains by introducing local linear sources $j(x)$ for $\chi(x)$ and performing a Legendre transform of $\ln Z[j]$, with $\bar{\phi}(x)=\langle\chi(x)\rangle_{|j}$ in the presence of sources. The nonlocal sources $h(\vec{q})$ are not affected by this transformation and $\Gamma$ depends now on $h(\vec{q})$ and $\bar{\phi}(x)$. We also add a field-independent piece linear in $h(\vec{q})$ which takes care of the ``additive renormalization'' $\pm 1/2$ on the l.h.s.of eq. (\ref{18})
\begin{eqnarray}\label{19A}
\Gamma[\bar\phi]&=&-\ln Z[j]\\
&&+\int_q\big(\bar\phi^*(q)j(q)+j^*(q)\bar{\phi}(q)\big)\pm\frac{\Omega_{d+1}}{2}\int_{\vec q}h(\vec{q}).\nonumber
\end{eqnarray}
It is easy to verify that the occupation numbers obey
\begin{equation}\label{20}
n(\vec{q})=-\frac{1}{\Omega_{d+1}}\frac{\delta\Gamma}{\delta h(\vec{q})}_{|\bar{\phi}_0(x),h=0}.
\end{equation}
Here the partial derivative $\delta/\delta h(\vec{q})$ is taken at fixed $\bar{\phi}(x)$ and evaluated for the solution of the field equation $\bar{\phi}_0(x)$ which obeys
\begin{equation}\label{21}
\frac{\delta\Gamma}{\delta\bar{\phi}(x)}_{|\bar{\phi}_0(x),h=0}=0.
\end{equation}

Our goal will be the computation of $\Gamma[\bar{\phi};h]$ in linear order in $h$. Expanding 
\begin{equation}\label{22}
\Gamma[\bar{\phi};h]=\Gamma[\bar{\phi}]-\int_{\vec q}{\cal N}[\bar{\phi};\vec q]h(\vec{q})+0(h^2)
\end{equation}
the occupation numbers follow directly from ${\cal N}$
\begin{equation}\label{23}
n(\vec{q})=\frac{1}{\Omega_{d+1}}{\cal N}[\bar{\phi}_0;\vec q].
\end{equation}
The field equation for $\bar\phi_0$ only involves $\Gamma[\bar{\phi}]$.

\section{Exact flow equation for occupation numbers}
\label{flow}
The difference of ${\cal N}[\bar{\phi};\vec q]$ from its classical value results from quantum and thermal fluctuations. We include these fluctuation effects stepwise by introducing first an infrared cutoff $R_k$ which suppresses the fluctuations with momenta $\vec{q}\ ^2<k^2$. (For fermions we may choose to suppress the fluctuations with $|\vec{q}\ ^2-k^2_F|<k^2.)$ At the end we lower $k$ to zero such that all fluctuations are included. We achieve this strategy by adding to the action (\ref{1}) an infrared cutoff term \cite{CWAV}. 
\begin{equation}\label{24}
\Delta_k S=\int_qR_k(\vec{q})\chi^*(q)\chi(q).
\end{equation}
In turn, the effective action is now replaced by the average action $\Gamma_k$ which depends on $k$ \cite{CWFE}, \cite{BTW}. With $R_k(\vec{q})$ diverging for $k\to\infty$ all fluctuations are suppressed in this limit and one has $\Gamma_{k\to\infty}=S$. On the other hand $R_k(\vec{q})=0$ for $k\to 0$ implies $\Gamma_{k\to 0}=\Gamma$. The average action therefore interpolates smoothly between the classical action for $k\to \infty$ (or $k=\Lambda$) and the effective action for $k\to 0$. 

Its dependence on $k$ obeys an exact flow equation \cite{CWFE}
\begin{eqnarray}\label{25}
\partial_k\Gamma_k[\phi;h]=\frac12 {\rm Tr} \{\partial_k {\cal R}
(\Gamma^{(2)}_k[\phi;h]+{\cal R})^{-1}\},
\end{eqnarray}
with ${\cal R}(q,q')=R_k(\vec{q})\delta(q-q')$. The second functional derivative $\Gamma^{(2)}_k$ is given by the full inverse propagator in the presence of ``background fields'' $\bar{\phi}$. For a homogeneous background field one has $\Gamma^{(2)}_k(q,q')=\bar P(q)\delta(q-q')$, with $\bar P$ a matrix in the space of complex fields $(\bar{\phi},\bar{\phi}^*)$ or real fields $(\bar{\phi}_1,\bar{\phi}_2)$, with,  $\bar{\phi}=\frac{1}{\sqrt{2}}(\bar{\phi}_1+i\bar{\phi}_2)$. The trace involves a momentum integration and trace over internal indices. We will choose here the basis $(\bar{\phi}_1,\bar{\phi}_2)$. For homogeneous background fields $\bar{\phi}$ eq. (\ref{25}) takes the explicit form (with $tr$ the internal trace)
\begin{equation}\label{26}
\partial_k\Gamma_k=\frac{\Omega_{d+1}}{2}tr\int_q\partial_kR_k(\vec{q})(\gamma_2(q)+R_k(\vec{q})^{-1}.
\end{equation}
The precise shape of the cutoff function $R_k$ is, in principle, arbitrary. As an example, one may consider the ``optimized cutoff'' \cite{Litim}
\begin{equation}\label{27}
R_k(\vec{q})=\frac{\bar A}{2M}(k^2-\vec{q}\ ^2)\Theta(k^2-\vec{q}\ ^2).
\end{equation}
We observe that for $R_k$ depending only on $\vec q$ the nonlocal source term (\ref{17}) may be viewed as a shift $R_k(\vec{q})\to R_k(\vec{q})-h(\vec{q})$ such that a variation with respect to $h$ can also be interpreted as a variation in the space of cutoff functions. 

The presence of the nonlocal source $h(\vec{q})$ does not change the general structure of the exact flow equation (\ref{25}). Employing again the expansion (\ref{22}) (now with all quantities depending on $k$) the flow for $\Gamma_k[\bar\phi]$ obtains from eq. (\ref{25}) by simply omitting the $h$-dependence, corresponding an evaluation for $h=0$. The solution of the field equation (\ref{21}) $\bar{\phi}_0(k)$ will now depend on $k$. The exact flow equation for ${\cal N}_k[\bar{\phi};\vec q]$ obtains from eq. (\ref{25}) by taking a derivative with respect to $h(\vec{q})$
\begin{equation}\label{28}
\partial_k{\cal N}_k=-\frac12 {\rm Tr} \{\partial_k {\cal R}(\Gamma^{(2)}_k+{\cal R})^{-1}{\cal N}^{(2)}_k(\Gamma^{(2)}_k+{\cal R})^{-1}\}.
\end{equation}
Here ${\cal N}^{(2)}_k(q',q'')$ is the second functional derivative
\begin{equation}\label{29}
{\cal N}^{(2)}_{k,ab}(q',q'')=
\frac{\delta^2{\cal N}_k}{\delta\bar{\phi}^*_a(q')\delta\bar{\phi}_b(q'')}.
\end{equation}
Eq. (\ref{28}) will be our basic starting point for a computation of $n(\vec{q})$ through a flow equation. We will consider a $k$-dependent $n_k(\vec{q})$ as defined by eq. (\ref{23}) and extract the occupation numbers for $k\to 0$.

For a homogenous $\bar\phi_0(k)$ one has ${\cal N}^{(2)}_k(q',q'')=\nu_2(q')\delta(q'-q'')$ and the flow equation for ${\cal N}_k$ simplifies
\begin{eqnarray}\label{32}
&&\partial_k{\cal N}_k[\bar{\phi}_0]=\\
&&-\frac{\Omega_{d+1}}{2}tr
\int_{q'}\partial_kR_k(\vec q\ ')\nu_2(q')[\bar P(q')+R_k(\vec q\ ')]^{-2}\nonumber
\end{eqnarray}
where $\nu_2$ is evaluated for $\bar{\phi}=\bar{\phi}_0$. This results in the final flow equation for the occupation numbers
\begin{eqnarray}\label{33}
&&\partial_k n_k(\vec{q})=-\frac12 tr\int_{q'}\partial_kR_k(\vec q\ ')
\nu_2(q';\vec q)\nonumber\\
&&[\bar P(q')
+R_k(\vec q\ ')]^{-2}+\frac{\partial n_k(\vec{q})}{\partial\bar\phi_0^2}\partial_k\bar\phi_0^2.
\end{eqnarray}
Here the second term accounts for the $k$-dependence of the solution $\bar\phi_0(k)$.

For $k\to \infty$ the fluctuation effects can be computed exactly. For bosons and $\sigma\leq 0$ we start for $k=\Lambda$ with the initial value
\begin{equation}\label{30}
{\cal N}_\Lambda[\bar{\phi};\vec q]=\frac12\sum\limits_a\int_{q_0}\bar{\phi}^*_a(q_0,\vec q)\bar{\phi}_a(q_0,\vec q)
\end{equation}
and
\begin{equation}\label{31}
{\cal N}^{(2)}_\Lambda[\bar{\phi};\vec q](q',q'')=\delta(\vec q-\vec q\ ')\delta(q'-q'')\delta_{ab}.
\end{equation}
More precisely, the initial value for $k\to\infty$ has a classical term and a one loop contribution. For bosons and $\sigma\leq 0$ the latter precisely cancels the classical contribution from the last term in eq. (\ref{19A}). Since there is essentially no running between $k\to\infty$ and $k=\Lambda$ we may begin the flow at $k=\Lambda$. (For fermions and $\sigma>0$ the one loop effect is less simple, see eq. (\ref{59}) below.) According to eq. (\ref{30}) one has for $\sigma\leq0$ the initial value 
\begin{equation}\label{31A}
n_\Lambda(\vec{q})=\bar{\phi}^2_0(\Lambda)\delta(\vec{q}).
\end{equation}
The only possible contribution arises from a condensate $\bar\phi_0(\Lambda)\neq 0$. For $\vec q\neq 0$ a nonzero occupation number has to be built up as a result of the flow. 

We want to extract $n(\vec{q})$ from a solution of the flow equation (\ref{28}) for $k\to0$. This will only be possible in suitable approximations. One may imagine to solve first the flow equation for $\Gamma_k$ and then use the result in eq. (\ref{29}). For this purpose we will need $\nu_2(q';\vec q)$. 

The quantity $\nu$ obeys a type of Ward identity. We have defined $n_p(\vec q)$ in terms of the $k$-dependent propagator.  Using the exact flow equation for the exact propagator \cite{BTW} yields a second exact equation for the flow of $n_p$. The comparison of the two exact flow equations yields the ``Ward identity''.

Let is demonstrate this for a homogeneous setting where
\be\label{89A}
n_{p,k}(\vec q)=\frac12\int_{q_0}tr\bar G(q_0,\vec q)-\frac12,
\ee
with $\bar G$ the $k$-dependent propagator matrix. For $\vec q\neq 0$ this yields the flow equation
\be\label{89B}
\partial_kn_k(\vec q)=\zeta_n(\vec q)+\frac{\partial n_k(\vec q)}{\partial\bar\phi^2_0}
\partial_k\bar\phi^2_0,
\ee
with
\ba\label{89C}
\zeta_n(\vec q)&=&\frac12\int_{q_0}tr\partial_k\bar G(q_0,\vec q){×}\\
&=&-\frac12\int_{q_0}tr\big[(\bar P+R_k)^{-2}\partial_k(\bar P+R_k)\big](q_0,\vec q).\nn
\ea
If we decompose
\be\label{89D}
\nu (q';\vec q)=\delta(\vec q\ '-\vec q)+\nu_p(q',\vec q)
\ee
we obtain by comparing eqs. \eqref{33} and \eqref{89B}
\be\label{89E}
tr\int_{q'}\partial_kR_k(q')\nu_p(q';\vec q)\bar G^2(q')=tr\int_{q_0}\partial_k\bar P(q_0,\vec q)
\bar G^2(q_0,\vec q).
\ee
With $\Gamma^{(2)}(q,q')=\bar P(q)\delta(q-q')$ the exact flow equation for $\bar P$ follows from the second functional derivative of eq. \eqref{25} \cite{CWFE}. It can be written as a one loop expression
\be\label{89F}
\partial_k\bar P(q)=tr\int_{q'}\partial_kR_k(q')\bar G^2(q') {\cal H}(q';q),
\ee
with ${\cal H}$ a matrix both in the space where $\bar G$ acts (and over which the trace in eq. \eqref{89F} is taken) as well as in the space where $\bar P$ acts. We identify
\be\label{89G}
\nu_p(q';\vec q)=\tilde{tr} \int_{q_0}{\cal H}(q';q_0,\vec q)\bar G^2(q_0,\vec q),
\ee
with $\tilde{tr}$ denoting the trace in the space where $\bar G$ acts. 

\section{Quadratic truncation}
\label{quadratic}
The flow equation (\ref{33}) is exact. However, even if $\bar P(q')$ would be known from a solution to the flow equation for $\Gamma_k[\bar{\phi}]$, the r.h.s. still involves the unknown second functional derivative $\nu_2$. One may derive an exact flow equation for $\nu_2$, but in turn it will involve even higher functional derivatives of ${\cal N}$. We need to approximate the solution by a suitable ansatz (truncation) for the general form of ${\cal N}$.

As a first approach we consider bosons and truncate ${\cal N}$ in the number of fields. We keep only terms with up to two powers of $\bar\phi$
\begin{eqnarray}\label{35}
{\cal N}_k[\bar{\phi};\vec q]&=&\Omega_{d+1}n_{p,k}(\vec{q}) \nonumber\\
&&+\frac12\sum\limits_a\Big\{
\int_{q'_0}Z_{c,k}(q'_0,\vec q)\bar{\phi}^\dagger_a(q'_0,\vec q)\bar{\phi}_a(q'_0,\vec q)\nonumber\\
&&+\int_{q'} Z_{p,k}(q',\vec q)\bar{\phi}^\dagger_a(q')\bar{\phi}_a(q')\Big\}\nonumber\\
&&-\Omega_{d+1} Z_{p,k}(0,\vec q)\bar{\phi}_0^2.
\end{eqnarray}
The initial values are
\begin{equation}\label{36}
n_{p,\Lambda}(\vec{q})=0~,~Z_{c,\Lambda}(q'_0,\vec q)=1~,~Z_{p,\Lambda}(q',\vec q)=0.
\end{equation}
For $\bar{\phi}_0(\vec{q})=\bar{\phi}_0\delta(q')$ the momentum distribution reads
\begin{equation}\label{37}
n(\vec{q})=n_{p,0}(\vec{q})+
Z_{c,0}(0,0)\bar{\phi}^2_0\delta(\vec{q}).
\end{equation}
The condensate density
\begin{equation}\label{38}
n_c=Z_{c,0}(0,0)\bar{\phi}_0^2
\end{equation}
differs from the bare order parameter $\phi_0^2$ only if $Z_c(0,0)\neq 1$. The total density $n$ and condensate fraction $\Omega_c$ read
\begin{equation}\label{39}
n=n_c+\int_{\vec q}n_p(\vec{q})~,~\Omega_c=n_c/n.
\end{equation}

We need to understand the meaning of the functions $Z_{c,k}$ and $Z_{p,k}$ and start by defining them properly. Let us first use a combined ``density transfer function'' $Z_h$ 
\begin{equation}\label{40}
Z_h(q',\vec q)=Z_p(q',\vec q)+Z_c(q'_0,\vec q)\delta(\vec q\ '-\vec q)
\end{equation}
and define $Z_p$ and $Z_c$ by their behavior for $\vec q\ '\to \vec q$. We want to choose $Z_p(q',\vec q)$ such that it remains finite for $\vec q\ '=\vec q$. We therefor define
\begin{equation}\label{40A}
Z_c(q'_0,\vec q)=\Omega^{-1}_d\int\limits_{\vec q\ '}~~Z_h(q',\vec q)\delta(\vec q\ '-\vec q),
\end{equation}
which yields $Z_p$ implicitely by subtraction in eq. (\ref{40}). (One may replace $\delta(\vec q\ '-\vec q)$ by a smoothened $\delta$-function and change $Z_p(q',\vec q)$ correspondingly. This concerns the practical definition which particles are counted as having zero momentum.) 

In terms of $Z_h$ the second functional derivative reads 
\begin{equation}\label{41}
{\cal N}^{(2)}(q',q'')=Z_h(q',\vec q)\delta(q'-q'')~,~\nu_2(q',\vec q)=Z_h(q',\vec q).
\end{equation}
All quantities need to be evaluated for the solution of the field equation $\bar{\phi}_0(k)$. The substraction of a constant term $\sim Z_{p,k}\bar\phi^2_0$ in the truncation (\ref{35}) guarantees that the expansion in powers of fields is correctly performed around $\bar\phi_0(k)$ and that $Z_p$ does not contribute to the particle density. 
The flow equation for $n(\vec{q})$ (\ref{33}) has both a contribution from $Z_p$ and $Z_c$
\begin{eqnarray}\label{42}
\partial_kn(\vec{q}) 
&=&-\frac12 tr\int_{q'_0}\partial_kR_k(\vec{q}) Z_c(q'_0,\vec q)\nonumber\\
&&\qquad\big[\bar P(q'_0,\vec q)+R_k(\vec{q})\big]^{-2}\nonumber\\
&&\hspace{-0.7cm}-\frac12 tr\int_{q'}\partial_kR_k(\vec q\ ')Z_p(q',\vec q)
\big[\bar P(q')+R_k(\vec q\ ')\big]^{-2}\nonumber\\
&&\hspace{-0.7cm}+\frac{\partial n(\vec{q})}{\partial\bar\phi_0^2}\partial_k\bar\phi_0^2.
\end{eqnarray}

In the remainder of this section we argue that the condensate density is precisely given by $\bar\phi^2_0$. This means that $Z_c(0,0)=1$ is not renormalized, as will be shown more precisely in the next section. We observe that both contributions to eq. \eqref{42} from $Z_p$ and $Z_c$ are regular for $\vec{q}\ ^2\to0$ and therefore do not contribute to the running of the condensate density $n_c$. The latter simply obeys for $\bar\phi_0\neq 0$
\begin{equation}\label{43}
\partial_kn_c=\frac{\partial n_c}{\partial\bar\phi_0^2}\partial_k\bar\phi_0^2.
\end{equation}
and $\partial_kn_c=0$ for $\bar\phi_0=0$. 

We will still need the $\bar\phi_0$-dependence of $n(\vec{q})$. Again, we derive a flow equation for
\begin{equation}\label{43A}
\alpha_k(\vec{q})=\frac{\partial n_k(\vec{q})}{\partial\bar\phi_0^2}~,~
\alpha_\Lambda(\vec{q})=\delta(\vec{q}).
\end{equation}
It reads $({\cal G}^{-1}=\Gamma^{(2)}+{\cal R})$
\begin{eqnarray}\label{43B}
&&\partial_k\alpha(\vec{q})=-
\frac{1}{2\Omega_{d+1}}{\rm Tr} 
\left\{\partial_k {\cal R}\frac{\partial}{\partial\bar\phi^2_0}({\cal G}{\cal N}^{(2)}{\cal G})\right\}\\
&&=\frac{1}{2}tr 
\int\limits_{q'}\left\{\bar G\partial_kR_k\bar G\left[\frac{\partial\Gamma^{(2)}}{\partial\bar\phi_0^2}\bar G Z_h\right.
\left.+Z_h\bar G\frac{\partial\Gamma^{(2)}}{\partial\bar\phi_0^2}\right]\right\}\nonumber
\end{eqnarray}
where we have omitted a term $\sim\partial Z_h/\partial\bar\phi_0^2$ in the last equation. The only dependence on $\vec q$ arises from $Z_h$ and we observe that the r.h.s. of eq. (\ref{43B}) remains regular for $\vec q\ ^2\to 0$. In consequence, we can use for all $k$ a non-singular $\tilde\alpha_k(\vec{q})$
\begin{equation}\label{43C}
\alpha(\vec{q})=\delta(\vec{q})+\tilde\alpha(\vec{q})
\end{equation}
and the condensate density obeys
\begin{equation}\label{43D}
\frac{\partial n_c}{\partial\bar\phi_0^2}=1~,~
\partial_k n_c=\partial_k\bar\phi_0^2~,~n_{c,k}=
\bar\phi^2_0(k).
\end{equation}
This closes the argument that the condensate density is not renormalized and shows that $n_p(\vec{q})$ has no contribution $\sim\delta(\vec{q})$. For the SSB regime with $\bar\phi_0(k)^2>0$ we may define $Z_c(0,0)=n_c/\bar\phi_0^2$ such that
\begin{equation}\label{44}
\partial_kZ_c(0,0)=\frac{\partial Z_c(0,0)}{\partial\bar\phi_0^2}\partial_k\bar\phi_0^2=0.
\end{equation}
A simple solution with the initial condition of a $\bar\phi$ independent $Z_{c,\Lambda}(0,0)=1$ is $Z_c(0,0)=1$ for all $k$.

\section{Density transfer function}
\label{density}
The right hand side of the density flow equation (\ref{42}) involves the ``density transfer function'' $Z_h(q',\vec q)$ (\ref{40}). In order to understand its role we neglect for a moment the dependence of $Z_h$ on $q'_0$. The first two terms in eq. (\ref{42}) can then be written as
\begin{equation}\label{44A}
\partial_kn(\vec{q})=\int_{\vec q\ '}Z_h(\vec q\ ',\vec q)\partial_k\tilde n(\vec q\ ')+\dots
\end{equation}
where $\partial_t\tilde n(\vec{q})$ accounts for the contribution of a loop with momentum $\vec q\ '$
\begin{eqnarray}\label{44B}
\partial_k\tilde n(\vec q\ ')=-\frac12 tr\int_{q'_0}\partial_kR_k(\vec q\ ')\big(\bar P(q_0,\vec q\ ')
+R_k(\vec q\ ')\big)^{-2}.\nonumber\\
\end{eqnarray}
For a $q'_0$-dependent $Z_h$ the folding (\ref{44A}) is in a certain sense averaged over $q'_0$. The density transfer function describes how a loop with momentum $\vec q\ '$ influences the occupation number $n(\vec{q})$. 

The flow equation for $Z_h(q',\vec q)$ obtains by taking the second functional derivative of eq. (\ref{28}) with respect to $\bar\phi$. In the truncation (\ref{35}) ${\cal N}^{(2)}_k$ does not depend on $\bar\phi$ and one finds with ${\cal G}=(\Gamma^{(2)}_k+{\cal R})^{-1}$ 
\begin{eqnarray}\label{45}
&&\partial_k{\cal N}^{(2)}_{ab}(q',q'')=\frac12 Tr
\left\{[{\cal G}\partial_k {\cal R}{\cal G}{\cal N}^{(2)}{\cal G}\right.\nonumber\\
&&+{\cal G}{\cal N}^{(2)}{\cal G}\partial_k{\cal R}{\cal G}]
\frac{\delta^2\Gamma^{(2)}}{\delta\bar{\phi}^*_a(q')\delta\bar{\phi}_b(q'')}\Big\}\nonumber\\
&&-\frac12 {\rm Tr} \left\{{\cal G}\partial_k{\cal R}{\cal G}
\frac{\delta\Gamma^{(2)}}{\delta\bar\phi^*_a(q')}{\cal G}{\cal N}^{(2)}{\cal G}\frac{\delta\Gamma^{(2)}}{\delta\bar\phi_b(q'')}
\right\}\nonumber\\
&&-\frac12 {\rm Tr} \left\{{\cal G}\partial_k{\cal R}{\cal G}
\frac{\delta\Gamma^{(2)}}{\delta\bar\phi_b(q'')}{\cal G}{\cal N}^{(2)}{\cal G}\frac{\delta\Gamma^{(2)}}{\delta\bar\phi^*_a(q')}
\right\}\nonumber\\
&&-\frac12 \left\{{\rm Tr} {\cal G}({\cal N}^{(2)}{\cal G}\partial_k{\cal R}+\partial_k{\cal R}{\cal G}{\cal N}^{(2)}){\cal G}\right.\nonumber\\
&&\left.\left(
\frac{\partial\Gamma^{(2)}}{\partial\bar\phi^*_a(q')}{\cal G}
\frac{\partial\Gamma^{(2)}}{\partial\bar\phi_b(q'')}+
\frac{\partial\Gamma^{(2)}}{\partial\bar\phi_b(q'')}{\cal G}
\frac{\partial\Gamma^{(2)}}{\partial\bar\phi^*_a(q')}\right)\right\}.
\end{eqnarray}
The first term involves a quartic vertex and the two last terms arise from cubic vertices present for $\bar\phi_0\neq0$. We will approximate the vertices here by pointlike vertices
\begin{eqnarray}\label{46}
\frac{\delta\Gamma^{(2)}_{cd}(p',p'')}{\delta\bar{\phi}^*_a(q')\delta\bar{\phi}_b(q'')}&=&
\lambda_{abcd}\delta(p'-p''+q'-q''),\nonumber\\
\frac{\delta\Gamma^{(2)}_{cd}(p',p'')}{\delta\bar\phi^*_a(q')}&=&\gamma_{acd}\bar\phi_0\delta(p'-p''+q'),\nonumber\\
\frac{\delta\Gamma^{(2)}_{cd}(p',p'')}{\delta\phi_b(q'')}&=&\bar\gamma_{bcd}\bar\phi_0\delta(p'-p''-q'').
\end{eqnarray}
For a quartic coupling $\lambda_\phi$ one has
\begin{eqnarray}\label{46A}
\gamma_{acd}&=&\bar\gamma_{acd}=\sqrt{2}\lambda(\delta_{a1}\delta_{cd}+\delta_{c1}\delta_{ad}+\delta_{d1}\delta_{ac}),\nonumber\\
\lambda_{abcd}&=&\lambda(\delta_{ab}\delta_{cd}+\delta_{ac}\delta_{bd}+\delta_{ad}\delta_{bc}).
\end{eqnarray}
For a homogeneous setting ${\cal G}$ is diagonal in momentum space. One obtains the flow equations
\begin{eqnarray}\label{47}
\partial_k{\cal N}^{(2)}_{cc}&&\hspace{-0.3cm}(q',q'';\vec q)=
\delta(q'-q'')\sum\limits^2_{a,b=1}
\int\limits_{p'}\partial_kR_k(\vec p')\nonumber\\
&&\Big\{\lambda_{ccab}(\bar{G}^3)_{ba}(p')Z_h(p',\vec q)\nonumber\\
&&-\frac{\bar\phi_0^2}{2}\sum\limits^2_{d,e=1}
\Big[(\bar{G}^2)_{ab}(p')\nonumber\\
&&\big(\gamma_{cbd}\bar\gamma_{cea}(\bar{G}^2)_{de}(p'+q')Z_h(p'+q',\vec q)\nonumber\\
&&+\bar\gamma_{cbd}\gamma_{cea}(\bar{G}^2)_{de}(p'-q')Z_h(p'-q',\vec q)\big)\nonumber\\
&&+2(\bar{G}^3)_{ab}(p')Z_h(p',\vec q)(\gamma_{cbd}\bar\gamma_{cea}\bar{G}_{de}(p'+q')\nonumber\\
&&+\bar\gamma_{cbd}\gamma_{cea}\bar{G}_{de}(p'-q')\big)\Big]\Big\}
\end{eqnarray}
Since $\partial_k{\cal N}^{(2)}_{11}$ gets, in principle, also contributions from higher order terms neglected in our truncation we identify
\begin{equation}\label{48}
\partial_kZ_h(q',\vec q)\delta(q'-q'')=\partial_k{\cal N}^{(2)}_{22}(q',q'';\vec q).
\end{equation}

The r.h.s. of eq. (\ref{47}) remains finite for $\bar q'\to\vec q$. The contribution from $Z_c$ changes the $q'$-integration to an integration over $q'_0$ with values of $\vec p'$ fixed by the $\delta$-function in $Z_h$, i.e. $\vec p'=\vec q$ or $\vec p'=\vec q\pm\vec q\ '$. The $q_0$-integration is well behaved due to the infrared cutoff $R_k$ in $\bar G^{-1}$. This leads to the important conclusion that renormalization effects for the condensate contribution are absent
\begin{equation}\label{49}
\partial_kZ_c(q'_0,\vec q)=0.
\end{equation}
The condensate density is therefore given by the ``bare'' order parameter
\begin{equation}\label{50}
n_c=\bar{\phi}_0^2.
\end{equation}
This is consistent with the discussion after eq. (\ref{44}).

In contrast, the build up of the occupation number for uncondensed particles is influenced by the renormalization of $Z_p$
\begin{eqnarray}\label{51}
&&\partial_kn_p(\vec{q})=\nonumber\\
&&-\frac12 tr \int_{q'_0}\partial_kR_k(\vec q)[\bar P(q'_0,\vec q)+R_k(\vec q)]^{-2}\nonumber\\
&&-\frac12 tr \int_{q'}\partial_kR_k(\vec q\ ')Z_p(q',\vec q)[\bar P(q')+R_k(\vec q\ ')]^{-2}\nonumber\\
&&+\frac{\partial{n_p}(\vec{q})}{\partial\bar\phi_0^2}\partial_k\bar\phi_0^2.
\end{eqnarray}
Only the first term in eq. (\ref{51}) is present in the ``classical contribution'' that neglects the renormalization effects for $Z_h$. For the second term the initial value $Z_p=0$ does not remain a solution of the flow equation. A nonzero $Z_p$ is generated by the contribution $\delta(\vec q\ '-\vec q)$ in $Z_h(q',\vec q)$ in the r.h.s of eq. (\ref{47}). Whereas the first term involves only ``on shell momenta'' where the propagator in the loop is fixed to $\vec q\ '=\vec q$, the second term gets also contributions from ``off shell momenta'', i.e. from $\vec q\ '$ different from $\vec q$. This reflects the broadening of the density transfer function $Z_h(q',\vec q)$ due to the renormalization flow. 

\section{Occupation number flow at low temperature}
\label{densityflow}
In order to get some practice with the flow equations  we first check simple situations. We start with the vacuum. The term $\sim\phi_0^2$ in eq. (\ref{47}) is absent. We want to compute the contribution from $Z_c$ to the flow of $Z_p$
\begin{equation}\label{52}
\partial_kZ_p(p',\vec q)=\sum\limits^2_{a=1}\lambda_{22aa}\partial_kR_k(\vec{q})\int_{q'_0}(\bar{G}^3)_{aa}(q'_0,\vec q).
\end{equation}
It is convenient to work in the complex basis with diagonal $\bar{G}$. For large $|q_0|$ the integrand decays $\sim |q_0|^{-3}$ such that we can close the integral either on the upper or lower half plane. The integral vanishes, as most easily seen in the complex basis where $\bar{G}$ has only one pole in the upper half plane and we can evaluate the closed integral in the lower half plane. In consequence, $Z_p$ stays zero for all $k$ and the density transfer function remains a $\delta$-function
\begin{equation}\label{53}
Z_h(p',\vec q)=\delta(\vec p'-\vec q).
\end{equation}
Inserting this result in eq. (\ref{51}) only the first term contributes
\begin{equation}\label{54}
\partial_kn_p(\vec{q})=-\frac12\partial_kR_k(\vec{q}) tr\int_{q'_0}\bar{G}^2(q'_0,\vec q).
\end{equation}
Again, $\bar{G}^2(q'_0,\vec q)$ decays sufficiently fast for large $|q'_0|$ and the r.h.s. of eq. (\ref{54}) vanishes, as it should be.

Next we may have a look on the free theory for $T=0$. In the absence of interactions the density transfer function does not flow and eq. (\ref{54}) remains valid, with $\Gamma^{(2)}$ the classical inverse propagator. Let us use for $\sigma\leq 0$ the explicite cutoff (\ref{27}) such that
\begin{eqnarray}\label{55}
&&k\partial_kn(\vec{q})=-\frac{k^2}{M}\theta(k^2-\vec q\ ^2)\nonumber\\
&&\int_{q'_o}
\left(iq'_0+\frac{\vec q\ ^2}{2M}+\frac{k^2-\vec q\ ^2}{2M}\theta(k^2-\vec q\ ^2)-\sigma\right)^{-2}.
\end{eqnarray}
The occupation number $n(\vec{q})$ flows only as long as $k^2>\vec q\ ^2$, then the flow stops. In the range $k^2>\vec q\ ^2$ the flow equation simplifies
\begin{equation}\label{56}
\frac{\partial}{\partial k^2}n_p=-\frac{1}{2M}\int_{q'_0}\left(iq'_0+\frac{k^2}{2M}-\sigma\right)^{-2}=0
\end{equation}
Since for large $k=\Lambda$ the initial value $n_{p,\Lambda}(\vec{q})$ vanishes we conclude that only the codensate $n_c$ contributes to the particle number, $n=n_c$. 

For fermions a positive chemical potential $\sigma>0$ is relevant. Now eq. (\ref{55}) is ill defined when max $(\vec q\ ^2,k^2)=2M\sigma$. This reflects that the cutoff (\ref{27}) is not well adapted to the problem. Indeed, instead of cutting out the small momentum fluctuations an efficient cutoff should rather suppress the fluctuations close to the Fermi surface, i.e. with $|\vec q\ ^2-2M\sigma|<k^2$. For the fermions we will therefore employ the cutoff
\begin{eqnarray}\label{57}
R^{(F)}_k&=&\frac{1}{2M}(k^2sgn\ \xi-\xi)\theta(k^2-\xi)~,\nonumber\\
\xi&=&\vec q\ ^2-\theta(\sigma)\sigma.
\end{eqnarray}
For positive $\sigma$ the propagator in presence of $R^{(F)}_k$ becomes in the vicinity of the Fermi surface
\begin{equation}\label{58}
\bar{G}=\left[iq_0+\frac{k^2}{2M}sgn(\vec q\ ^2-2M\sigma)\right]^{-1},
\end{equation}
such that the pole in presence of $R^{(F)}_k$ remains in the same half plane in complex $q_0$-space as for $R^{(F)}_k=0$. As a result, the initial value of the flow is given by the Fermi distribution
\begin{equation}\label{59}
n_{F,\Lambda}(\vec{q})=\theta(2M\sigma-\vec q\ ^2).
\end{equation}
In turn, the flow equation (\ref{56}) is replaced by 
\begin{eqnarray}\label{60}
&&\frac{\partial}{\partial k^2}n_F(\vec{q})=\frac{sgn(\vec q\ ^2-2M\sigma)}{2M}\nonumber\\
&&\int_{q'_0}\left(iq'_0+\frac{k^2}{2M}sgn(\vec q\ ^2-2M\sigma)\right)^{-2}=0
\end{eqnarray}
and the Fermi distribution is maintained for $k\to0$. For $T>0$ the r.h.s. does not vanish and the flow accounts for the smoothening of the Fermi distribution.

We next turn on the interactions. The initial values $n_{p,\Lambda}(\vec{q})=0~,~n_{F,\Lambda}(\vec{q})=\theta(2M\sigma-\vec q\ ^2)$ are not changed. However, the flow of $n(\vec{q})$ can now account for nonvanishing $n_p(\vec{q})$ or for a smoothening of the Fermi surface due to fluctuations. There are two possible effects. The first concerns the modifications of the propagator. Furthermore, a second effect arises from the build up of a nonvanishing $Z_p$. 

Let us first consider bosons and employ the cutoff (\ref{27}). The flow of $\tilde n(\vec{q})$ (cf. eq. (\ref{44B}) becomes $(\eta=-k\partial_k\ln\bar A)$
\begin{eqnarray}\label{61}
&&\partial_k\tilde n(\vec{q})=-\frac{k\bar A}{2M}\theta(k^2-\vec q\ ^2)\left(1-\frac\eta 2\right)\nonumber\\
&&\hspace{1.2cm} \int_{q'_0}(\bar{G}^2_{11}+\bar{G}^2_{22}+2\bar{G}_{12}\bar{G}_{21}).
\end{eqnarray}
We use the general form (\ref{16D})
\begin{equation}\label{62}
\bar{G}^2_{11}+\bar{G}^2_{22}+2\bar G_{12}\bar{G}_{21}
=\frac{(a+R_k)^2+(b+R_k)^2-2q^2_0Z_\phi^2}{\big[(a+R_k)(b+R_k)+q^2_0Z_\phi^2\big]^2}
\end{equation}
with momentum dependent coefficients $a,b,Z_\phi$. In order to understand the general structure we concentrate on large $\vec q\ ^2$ where perturbation theory should remain valid. We want to see in which limit we recover the Bogoliubov theory and what are the extensions of it. For this purpose we may and employ the ansatz (\ref{16J}). Neglecting $\eta$ and a possible $q_0$-dependence of $\bar A,\lambda_\phi,Z_\phi$ this yields $(Z_\phi=\bar AS)$
\begin{eqnarray}\label{63}
&&\partial_k\tilde n(\vec{q})=-\frac{k}{2M\bar A}\theta(k^2-\vec q\ ^2)\int_{q_0}\nonumber\\
&&\frac{(k^2/2M+2\lambda_\phi\bar\phi_0^2)^2+(k^2/2M)^2-2q^2_0S^2}
{\big[(k^2/2M)(k^2/2M+2\lambda_\phi\bar\phi^2_0)+q^2_0S^2\big]^2}\nonumber\\
&&=-\frac{4M^2\lambda^2_\phi\bar\phi_0^4}{Z_\phi k^2(k^2+4M\lambda_\phi\bar\phi_0^2)^{3/2}}
\theta(k^2-\vec q\ ^2).
\end{eqnarray}

For $k$-independent $\lambda_\phi,\bar\phi_0^2$ and $Z_\phi$ the integration of eq. (\ref{63}) yields 
\ba\label{127N}
\tilde n(\vec q)&=&\int\limits^\Lambda_q dk
\frac{4M^2\lambda^2_\phi\bar\phi_0^4}{Z_\phi k^2(k^2+4M\lambda_\phi\bar\phi^2_0)^{3/2}}
+\tilde n_\Lambda(\vec q)\\
&=&\frac{1}{2Z_\phi}\left[\left(1+
\frac{4M^2\lambda^2_\phi\bar\phi_0^4}{q^2(q^2+4M\lambda_\phi\bar\phi^2_0)}\right)^{1/2}-1\right]\nn\\
&&-\{q\to\Lambda\}+\tilde n_\Lambda(\vec q)\nn
\ea
For $\Lambda\to\infty,~\tilde n_\Lambda(\vec q)=0$ (we only consider here $\vec q\neq 0)$ the sum of the last two terms vanishes, such that
\be\label{127K}
\tilde n(\vec q)=\frac{1}{2Z_\phi}
\left[\left(1+
\frac{4M^2\lambda^2_\phi\bar\phi_0^4}{q^2(q^2+4M\lambda_\phi\bar\phi^2_0)}\right)^{1/2}-1\right].
\ee
If we replace $\tilde n$ by $n$ this is almost eq. \eqref{16K} - the difference vanishes for $Z_\phi=1$. The Bogoliubov result therefore obtains if we use the ``classical transfer function'' $Z_c=1,~Z_p=0$ and neglect the dependence of $a,b,Z_\phi$ on $q_0,\vec q$ and $k$.

Extensions beyond the Bogoliubov theory take into account the running and the momentum dependence of the parameters as well as a non-trivial transfer function. In a first step we may use in eq. \eqref{127K} effective parameters $\lambda_\phi, \bar\phi_0,Z_\phi$ which depend on $\vec q\ ^2$. Based on an approximate solution of the flow equation for the occupation numbers we argue that one should use the values $\lambda_\phi(\vec q,k^2=\vec q\ ^2)$ and $\bar\phi_0(k^2=\vec q\ ^2)$ instead of the values at $k=0$. This is justified since the build up of a nonvanishing $\tilde n(\vec{q})$ is dominated by the range $k^2\approx\vec q\ ^2$. 

Indeed, the flow \eqref{63}  stops for $k^2<\vec q\ ^2$, while for $k^2\gg\vec q\ ^2$ it is suppressed by inverse powers of $k$. Using also $Z_\phi=Z_\phi(\vec q,k^2=q^2)$ we note that $Z_\phi$ multiplies now a term that vanishes for $q\to\infty$, in distinction from eq. \eqref{16K}. As already noted in sect. \ref{improvement} this renormalization group improvement gets rid of the dangerous term $\sim \Delta Z_\phi$ in eq. \eqref{N11}. Indeed, the replacement of $Z_\phi(k)$ by a $k$-independent constant is only allowed if $Z_\phi$ multiplies a quantity that vanishes for large $q$, as in eq. \eqref{127K}. This is necessary since $Z_\phi$ multiplies in the flow equation \eqref{63} a term $\sim k^{-5}$ for large $k$. The properly implemented renormalization group flow takes care of the ultraviolet problems, even if the resolution of the $q_0$-dependence of the propagator is only poor. This has direct consequences for the $\Lambda$-dependence of the relation between particle density and chemical potential, $n(\sigma)$. For large $\Lambda$ the momentum integration (for $q<\Lambda)$ of $n(\vec q)$ yields $n\sim \Lambda^{d-4}$, with a coefficient that depends on $\sigma$. For $T=0$ the cutoff dependence of $n(\sigma)$ is suppressed by $(2M\sigma/\Lambda^2)^{(4-d)/2}$. 

A second improvement beyond Bogoliubov theory becomes necessary if we consider small $q$. Now the flow of the occupation numbers should also contain the term $\sim V q^2_0$ in the inverse propagator, as we have discussed in sect. \ref{Bosonoccupationnumbers}. 

Let us finally discuss the flow of the density transfer function which generates a nonvanishing $Z_p$ according to eq. (\ref{47}). The cubic couplings must have an even number of indices taking the value two. They are given by
\begin{eqnarray}\label{65}
&&\gamma_{212}=\gamma_{221}=\gamma_{122}=\bar\gamma_{212}=\bar\gamma_{221}=
\bar\gamma_{122}=\sqrt{2}\lambda_\phi,\nonumber\\
&&\gamma_{111}=\bar\gamma_{111}=3\sqrt{2}\lambda_\phi.
\end{eqnarray}
and the flow induced by these cubic couplings becomes 
\begin{eqnarray}\label{66}
&&\partial_kZ_{p,3}(q',\vec q)=-\frac{\lambda^2_\phi k\bar\phi_0^2}{M}\int_{p'}\theta(k^2-\vec p^{'2})\nonumber\\
&&\Big\{\big[(\bar{G}^2)_{11}(p')(\bar{G}^2)_{22}(p'+q')+(\bar{G}^2)_{22}(p')(\bar{G}^2)_{11}(p'+q')\nonumber\\
&&+\bar{G}^2_{12}(p')(\bar{G}^2)_{12}(p'+q')+(\bar{G}^2)_{21}(p')(\bar{G}^2)_{21}(p'+q')\big]\nonumber\\
&&\hspace{1.5cm}Z_h(p'+q',\vec q)\nonumber\\
&&+\big[(\bar{G}^3)_{11}(p')\bar{G}_{22}(p'+q')+(\bar{G}^3)_{22}(p')\bar{G}_{11}(p'+q')\nonumber\\
&&+(\bar{G}^3)_{12}(p')\bar{G}_{12}(p'+q')+(\bar{G}^3)_{21}(p')\bar{G}_{21}(p'+q')\big]\nonumber\\
&&\hspace{1.5cm}Z_h(p',\vec q)+(q'\to-q')\Big\}.
\end{eqnarray}
We evaluate eq. (\ref{66}) for $q'=0$, with the ansatz (\ref{16D}) and $\tilde a=a+R_k~,~\tilde b=b+R_k$, 
\begin{eqnarray}\label{67}
\partial_k Z_{p,3}(0,\vec q)&=&- 4\lambda^2_\phi k\bar\phi_0^2\int_{p'}\theta(k^2-\vec p^{'2})Z_h(p',\vec q)\nonumber\\
&&(\tilde a+\tilde b)^2(\tilde a\tilde b+p'^{2}_0Z_\phi^2)^{-3}.
\end{eqnarray}

For the high momentum region we use eq. (\ref{16J}) and the cutoff $R_k$ (\ref{B2}) such that
\begin{equation}\label{68}
\tilde b=\frac{\bar A}{2M}k^2~,~\tilde a=\tilde b+2\bar A\lambda_\phi\bar\phi_0^2.
\end{equation}
The flow
\begin{eqnarray}\label{69}
&&\hspace{-1.0cm}\partial_kZ_{p,3}(0,\vec q)=\nonumber\\
&&-\frac{12M^2\lambda^2_\phi\bar\phi_0^2}
{Z_\phi \bar A^2k^4}\frac{(k^2+2M\lambda_\phi\bar\phi_0^2)^2}{(k^2+4M\lambda_\phi\bar\phi_0^2)^{5/2}}\nonumber\\
&&\Big\{\theta(k^2-\vec q\ ^2)+\int_{\vec p}\theta(k^2-\vec p^2)Z_p(\vec p,\vec q)\Big\}
\end{eqnarray}
stops for $\vec q\ ^2<k^2$ and is dominated by $k^2\approx \vec q\ ^2$. As long as $Z_p$ remains small we can neglect the second term in the curled bracket. Omitting also the contribution from the four point vertex one finds for large $q$
\be\label{133}
Z_p(0,\vec q)=\frac{3M^2\lambda^2_\phi\bar\phi^2_0}{Z_\phi\bar A^2q^4}.
\ee

We conclude that $Z_p$ appears only in higher order in perturbation theory ($\sim\lambda^2_\phi)$ and is suppressed for large $q$. For a discussion of the ultraviolet behavior it can be neglected. For lower $q$, however, it may become necessary to include a nontrivial transfer function $Z_p$ if one aims for quantitative accuracy. It may be evaluated alternatively from eq. \eqref{89F}. 

\section{Gaps and pseudogaps}
\label{gapsandpseudo}
Besides a formal theoretical framework for systematic non-perturbative approximations, the exact flow equation for the occupation numbers can also be employed to motivate comparable simply renormalization group improvements. Examples are the discussion in sect. \ref{improvement} or the computation of the occupation numbers by a solution of the differential equation \eqref{68X}, with parameters evaluated at $k^2=q^2$. In this section we discuss gaps and pseudogaps in the light of such renormalization group improvements.

Let us first discuss fermions in presence of an order parameter or mean field. For definiteness, we consider a system of fermionic atoms, interacting with an effective bosonic di-atom fields $\bar\phi$. This field can represent microscopic molecules as well as Cooper pairs or other collective states with atom number two. For our purpose it is sufficient to consider a rather simple effective action for fermion fields $\psi$ with two ``spin states'' and $\bar\phi$
\ba\label{T1}
\Gamma&=&\int_x\left\{\psi^\dagger\left(\partial_\tau-\frac{\Delta}{2M}-\sigma\right)\psi\right.\nn\\
&&\left.-\frac{\bar h_\phi}{2}(\bar\phi^*\psi^T\epsilon\psi-\bar\phi\psi^\dagger\epsilon\psi^*)\right\}.
\ea
The Yukawa or Feshbach coupling $\bar h_\phi$ couples the fermions to bosons. For a constant real order parameter $\bar\phi(x)=\bar\phi_0$ this action is Gaussian and reads in momentum space
\be\label{T2}
\Gamma=\frac{1}{2}\int_q\tilde\psi^T(-q){\cal P}_F(q)\tilde\psi(q)~,~
\tilde\psi(q)=\left(\begin{array}{l}
\psi(q)\\\psi^*(-q)
\end{array}\right).
\ee
The inverse propagator ${\cal P}_F$ is a $4\times 4$ matrix \cite{DW}. 
\ba\label{S1}
{\cal P}_F(q)=\left(\begin{array}{ccc}
-\bar h_\phi\bar\phi_0\epsilon_{\alpha\beta}&,&(iq_0-\frac{\vec q\ ^2}{2M}+\sigma)\delta_{\alpha\beta}\\
(iq_0+\frac{\vec q\ ^2}{2M}-\sigma)\delta_{\alpha\beta}&,&\bar h_\phi\bar\phi_0\epsilon_{\alpha\beta}
\end{array}\right),\nn\\
\ea
where $\alpha,\beta=1,2$ are spin indices and $\epsilon_{12}=-\epsilon_{21}=1$. Correspondingly, the propagator matrix reads
\ba\label{S2}
&&\det\nolimits_F{\cal P}^{-1}_F=\\
&&\left(\begin{array}{ccc}
\bar h_\phi\bar\phi_0\epsilon_{\alpha\beta}&,&(-iq_0+\frac{\vec q\ ^2}{2M}-\sigma)\delta_{\alpha\beta}\\
(-iq_0-\frac{\vec q\ ^2}{2M}+\sigma)\delta_{\alpha\beta}&,&-\bar h_\phi\bar\phi_0\epsilon_{\alpha\beta}
\end{array}\right),\nn
\ea
with
\be\label{S3}
\det\nolimits_F=(iq_0+\frac{\vec q\ ^2}{2M}-\sigma)(-iq_0+\frac{\vec q\ ^2}{2M}-\sigma)+\bar h^2_\phi\bar\phi^2_0.
\ee
The fermionic occupation number obtains as 
\ba\label{S4}
n_F(\vec q)&=&-\frac12 \int_{q_0}tr
\left\{\left(\begin{array}{lll}
0&,&-\delta_{\alpha\beta}\\ \delta_{\alpha\beta}&,&0
\end{array}\right){\cal P}^{-1}_F\right\}+1\nn\\
&=&1-\int_{q_0}\left(\frac{q^2}{M^2}-2\sigma\right)\det\nolimits^{-1}_F.
\ea
For $\bar\phi_0=0$ we recover twice the result \eqref{9}, corresponding to the two spin states.

Let us evaluate $n_F(\vec q)$ for $T=0$. The zeros of $\det_F$ occur for 
\be\label{S5}
q^2_0=-\left\{\left(\frac{q^2}{2M}-\sigma\right)^2+\bar h^2_\phi\bar\phi^2_0\right\}.
\ee
One finds
\be\label{S6}
n_F(\vec q)=1-\left(\frac{q^2}{2M}-\sigma\right)
\left(\left(\frac{q^2}{2M}-\sigma\right)^2+\bar h^2_\phi\bar\phi^2_0\right)^{-1/2},
\ee
such that the Fermi surface is smoothened with a typical width set by the gap $\Delta=\bar h_\phi\bar\phi_0$. This generalizes to $T\neq 0$
\be\label{S7}
n_F(\vec q)=1-\frac{1}{\epsilon}
\left(\frac{q^2}{2M}-\sigma\right)+\frac{2}{\epsilon}\left(\frac{q^2}{2M}-\sigma\right)
\left[\exp\left(\frac{\epsilon}{T}\right)+1\right]^{-1},
\ee
with positive square root
\be\label{S8}
\epsilon=\sqrt{\left(\frac{q^2}{2M}-\sigma\right)^2+\Delta^2}\quad,\quad\Delta=\bar h_\phi\bar\phi_0.
\ee
In presence of a gap the fermion density $n_F$ does not vanish for $T=0$ and any finite $\sigma$. For $T=0$ and $\sigma=0$ the occupation numbers approach 
\be\label{S9}
n_F(\vec q)=1-
\frac{q^2}{\sqrt{q^4+4M^2\Delta^2}}\to 1-\frac{q^2}{2M\Delta}.
\ee
According to eq. \eqref{S5} the dispersion relation
\be\label{S10}
|\omega(\vec q)|=\epsilon(q)
\ee
is characterized by a minimal frequency set by the gap $|\omega|_{min}=\Delta$.

Renormalization group improvement implies that for any $\vec q$ the gap $\Delta (\vec q)$ should be evaluated at a characteristic momentum scale $k$. For fermions this implies
\be\label{S11}
\Delta(\vec q)=\bar h_\phi(k)\bar\phi_0(k)~,~
k^2=|\vec q\ ^2-2M\sigma|.
\ee
Effectively, the fermions with momenta distant from the Fermi surface, $\vec q=\vec q_F+\delta\vec q~,~\vec q_F\ ^2=2M\sigma$, are influenced by an effective volume with linear size $\sim k^{-1},$ where $k^2=|2\vec q_F\delta\vec q+\delta\vec q\ ^2|$. (The energy needed for a change $\delta\vec q$ in momentum is $k^2/(2M)$, and $k$ is a typical momentum of a boson whose absorption / emission can induce such a change.) The quantities $\bar h_\phi(k)$ and $\bar\phi_0(k)$ can be associated with the ``running couplings'' evaluated at the renormalization scale $k$. (This is precisely what happens if one investigates a $k$-dependent particle density as in \cite{DGPW}.) A non-vanishing gap $\Delta$ corresponds to
\be\label{S12}
\Delta=\bar h_\phi(k=0)\bar\phi_0(k=0)=h(k=0)\sqrt{\rho_0}
\ee
with renormalized Yukawa coupling $h=\bar h_\phi/\sqrt{\bar A}$. (We omit in our discussion the wave function renormalization for the fermions.) We observe that a gap does not necessarily require a nonzero order parameter $\bar\phi_0$. A nonvanishing renormalized order parameter $\sqrt{\rho_0}>0$ is sufficient, provided the renormalized Yukawa coupling remains also nonvanishing for $k\to 0$.

Furhermore, one may encounter situations where $\Delta(k)$ differs from zero for some range $k>k_{SR}$, while $\Delta(k<k_{SR})=0$. For small enough $k_{SR}$ this is a typical situation for a pseudogap. Except for a very narrow region around the Fermi surface the occupation numbers $n_F(\vec q)$ behave as if a gap were present. Such a situation has been discussed in detail for antiferromagnetic order in the two dimensional Hubbard model \cite{BBW}. A pseudogap occurs for temperatures between the critical temperature $T_c$ and the pseudocritical temperature $T_{pc}>T_c$. The presence of a pseudogap $\Delta(k)$ simply reflects the existence of local order in domains with linear size $\sim k^{-1}$. 

Similar considerations hold for bosonic occupation numbers. For non-relativistic bosons we may interprete $\Delta_r=\sqrt{2\lambda\rho_0}$ as a gap for the radial mode, while the Goldstone boson remains gapless. More complex gaps may occur if different bosonic degrees of freedom play a role. A gap for the bosons also exists in the symmetric phase where the effective potential $U(\rho)$ has a minimum at $\rho_0=0$, with $m^2=\partial U/\partial\rho_{|0}>0$. In this case one replaces in eq. \eqref{7A} $-\sigma\to m^2$. 

\section{Conclusions}
\label{conclusions}
In this paper we have concentrated on the more formal aspects of the functional integral approach to occupation numbers $n(\vec q)$ for interacting non-relativistic particles. Our study of bosons at zero temperature reveals pitfalls for computations beyond leading order perturbation theory - the Bogoliubov approximation. They are connected to the infrared and ultraviolet behavior of $n(\vec q)$. If insufficient care is taken, no meaningful result for the particle density $n$ will be obtained. This is a very practical problem for precision computations for ultracold atoms, where the relation between the particle density and the chemical potential $\sigma$ is a crucial quantity. One of the potential problems arises from an unphysical dependence of the relation between density and chemical potential, $n(\sigma)$, on the ultraviolet cutoff $\Lambda$.

A second class of problems concerns the infrared behavior of the occupation numbers. For $T=0$ we have found that the behavior of $n(\vec q)$ for $q\to 0$ differs qualitatively from the Bogoliubov theory for $d=1,2$ and mildly for $d=3$. This is due to the dominance of a term quadratic in the frequency in the full inverse propagator. Denoting with $V$ the corresponding coupling we find 
\ba\label{Z1}
&&n(\vec q)=\bar\phi^2_0\delta(\vec q)+n_p(\vec q),\nn\\
&&n_p(\vec q)\sim V^{-1/2}q^{-(1-\eta)}.
\ea
The anomalous dimension $\eta$ relates to the dense regime or Goldstone regime \cite{CWQP}. It is positive for $d=1$ and vanishes for $d=2,3$. As one of the consequences we find that the condensate depletion $n_p=\int_{\vec q}n_p(\vec q)$ is not dominated by the extreme infrared modes with $q\to 0$. This allows for a clear conceptual separation between the condensate density $n_c=\bar\phi^2_0$ and $n_p$. For $d=3$ the asymptotic behavior \eqref{Z1} sets in only for extremely small $q$, such that for most practical purposes the Bogoliubov approximation remains valid. 

For practical calculations with a good accuracy we propose to use the differential equation \eqref{68X} for the $q$-dependence of $n_p(\vec q)$. Correspondingly, $n_p$ can be computed from eq. \eqref{68Y}. The benefit of such a calculation is the absence of infrared or ultraviolet problems and the renormalization group improvement due to the use of running couplings. We postpone a numerical solution of eqs. \eqref{68X}, \eqref{68Y} to future work. 

We have derived an exact functional renormalization group equation for scale-dependent occupation numbers. They can be employed to motivate a renormalization group improvement in simpler settings. The effective couplings appearing in the evaluation of the occupation number $n(\vec q)$ for a particular momentum $\vec q$ correspond to running couplings, evaluated at a renormalization scale $k=q$ (with modifications in presence of a Fermi surface). This allows for an easy description of pseudogaps. 

\section*{APPENDIX A: One loop calculation of $Z_\phi$}
\renewcommand{\theequation}{A.\arabic{equation}}
\setcounter{equation}{0}
We define the wave function renormalization $Z_\phi(\omega,\vec q)$ by the $\omega$-derivative of the $2-1$-element of the inverse propagator
\begin{equation}\label{A.1}
Z_\phi(\omega,\vec q)=-i\frac{\partial}{\partial \omega}\bar P_{21}.
\end{equation}
For nonzero $\bar\phi_0$ the one loop contribution to $\bar P_{21}$ has a contribution from effective cubic couplings $\sim \lambda\bar\phi_0$
\begin{eqnarray}\label{A.2}
&&\Delta\bar P_{21}(\omega,\vec q)=-\lambda^2_\phi\bar\phi^2_0\int_{p_0}\int_{\vec p}\nonumber\\
&&\Big\{3\bar G_{11}(p_0,\vec p)
\big(\bar G_{21}(p_0+i\omega,\vec p+\vec q)+\bar G_{12}(p_0-i\omega,\vec p-\vec q)\big)\nonumber\\
&&+\bar G_{22}(p_0,\vec p)\big(\vec G_{12}(p_0+i\omega,\vec p+\vec q)+\bar G_{21}(p_0-i\omega,
\vec p-\vec q)\big)\Big\}\nonumber\\
&&=2i\omega\lambda^2_\phi\bar\phi^2_0\int_{p_0}\int_{\vec p}
\left(\frac{\vec p^2}{2M}-\lambda_\phi\bar\phi^2_0\right)\det\nolimits^{-1}(p)\nonumber\\
&& \qquad \qquad\quad\left(\det\nolimits^{-1}(p+q)+\det\nolimits^{-1}(p-q)\right),
\end{eqnarray}
with
\begin{equation}\label{A.3}
\det(p+q)=\frac{1}{4M^2}(\vec p+\vec q)^2
\big[(\vec p+q)^2+4M\lambda_\phi\bar\phi^2_0\big]+(p_0+i\omega)^2.
\end{equation}

We first consider the special case $\vec q=0$
\begin{eqnarray}\label{A.4}
Z_\phi(\omega,0)&=&2\lambda^2_\phi\bar\phi^2_0\int_{\vec p}
\left(\frac{\vec p^2}{2M}-\lambda_\phi\bar\phi^2_0\right)\nonumber\\
&&\hspace{-0.5cm}\frac{\partial}{\partial \omega}\Big\{\omega\big[
J(\vec p,\omega)+J(\vec p,-\omega)\big]\Big\},
\end{eqnarray}
where
\begin{eqnarray}\label{A.5}
J(\vec p,\omega)&=&\int_{p_0}\det\nolimits^{-1}(p)\det\nolimits^{-1}(p+q)\nonumber\\
&=&\int_{p_0}(p^2_0+B)^{-1}
\big((p_0+i\omega)^2+B\big)^{-1},\nonumber\\
B(\vec p)&=&\frac{1}{4M^2}\vec p^2
(\vec p^2+4M\lambda_\phi\bar\phi^2_0).
\end{eqnarray}
The integrand of the $p_0$-integration has simple poles at $p_0=\pm i\sqrt{B}$ and $p_0=\pm i\sqrt{B}-i\omega$ and we take $\omega\geq 0$ without loss of generality. The first pole in the upper half plane at $p_{0,1}=i\sqrt{B}$ has residuum
\begin{equation}\label{A.6}
r_1=\frac{i}{2\sqrt{B}\omega(2\sqrt{B}+\omega)},
\end{equation}
whereas a second pole in the upper half plane at $p_{0,2}=i(\sqrt{B}-\omega)$ exists only for $\omega<\sqrt{B}$, with residuum at 
\begin{equation}\label{A.7}
r_2=-\frac{i}{2\sqrt{B}\omega(2\sqrt{B}-\omega)}.
\end{equation}
This yields
\begin{equation}\label{A.8}
J(\omega>0)=-\frac{1}{2\sqrt{B}\omega}
\left(\frac{1}{2\sqrt{B}+\omega}-\frac{1}{2\sqrt{B}-\omega}\theta(\sqrt{B}-\omega)\right)
\end{equation}
or, for arbitrary $\omega$,
\begin{eqnarray}\label{A.9}
K(\omega)&=&\omega\big(J(\omega)+J(-\omega)\big)\nonumber\\
&=&\left\{
\begin{array}{cl}
\frac{2\omega}{\sqrt{B}(4B-\omega^2)}&{\rm for~}|\omega|<\sqrt{B}\\
-\frac{sign(\omega)}{\sqrt{B}(2\sqrt{B}+|\omega|)}&{\rm for~}|\omega|>\sqrt{B}
\end{array}
\right..
\end{eqnarray}
Both $K$ and $\partial K/\partial\omega$ are discontinuous at $|\omega|=\sqrt{B}$. 

At first sight the situation seems to simplify for $\omega=0$ where
\begin{equation}\label{A.10}
\frac{\partial K}{\partial \omega}_{|\omega=0}=\frac{1}{2B\sqrt{B}},
\end{equation}
resulting in 
\begin{equation}\label{A.11}
Z_\phi(0,0)=\lambda^2_\phi\bar\phi^2_0\int_{\vec p}
\left(\frac{\vec p^2}{2M}-\lambda_\phi\bar\phi^2_0\right)B^{-\frac32}(\vec p).
\end{equation}
With $\sigma=4M\lambda_\phi\bar\phi^2_0$ we have to solve the remaining integral over $y=|\vec p|$
\begin{eqnarray}\label{A.12}
Z_\phi(0,0)&=&4v_dM^2\lambda^2_\phi\bar\phi^2_0I(\sigma),\nonumber\\
I(\sigma)&=&\int\limits^\infty_0dy y^{d-4}
\left\{(y^2+\sigma)^{-1/2}-\frac{3\sigma}{2}(y^2+\sigma)^{-3/2}\right\}\nonumber\\
&=&\left(1+3\sigma\frac{\partial}{\partial\sigma}\right)\int\limits^\infty_0
dy y^{d-4}(y^2+\sigma)^{-1/2}\nonumber\\
&=&\left(\frac32 d-5\right)\sigma^{\frac{d-4}{2}}F_d,\nonumber\\
F_d&=&\int\limits^\infty_0dz z^{d-4}(z^2+1)^{-1/2}.
\end{eqnarray}
This yields
\begin{equation}\label{A.13}
Z_\phi(0,0)=-2^{d-3}(10-3d)v_d\\
F_d(\lambda_\phi M)^{\frac d2}\bar\phi_0^{d-2}.
\end{equation}

However, the integral $F_d$ is infrared finite only for $d>3$ and ultraviolet finite for $d<4$. For the realistic cases with $d\leq 3$ the limit $\omega\to 0$ has to be done more carefully. Even for very small $|\omega|$ eq. (\ref{A.10}) can only be used for $\sqrt{B}>|\omega|$ or 
\begin{equation}\label{A.14}
z>\frac{2M|\omega|}{\sigma}.
\end{equation}
This modifies effectively the factor $F_d$ in eq. \eqref{A.13}, which now becomes a function of $\omega/\sigma$. 
The contribution to $F_d$ from this range is approximately
\begin{equation}\label{A.15}
F^+_d=\left\{\begin{array}{cl}
\frac{1}{3-d}\left(\frac{2M\omega}{\sigma}\right)^{d-3}&{\rm for~}d<3\\
\ln\frac{\sigma}{2M\omega}&{\rm for~}d=3
\end{array}
\right.
\end{equation}
and diverges for $\omega\to 0$. In consequence, one has $Z_\phi(\omega,0)\sim\lambda^{\frac 32}_\phi\bar\phi_0$ for $d\leq 3$ with logarithmic corrections for $d=3$. The range $z<2M\omega/\sigma$ (for $\omega>0$) gives an infrared finite contribution for $d>1$ since for $z\to 0$ one has $\partial K/\partial\omega\approx 1/(\omega^2\sqrt{B})$ and the r.h.s. of eq. (\ref{A.11}) is multiplied by a factor $2B/\omega^2\approx z^2\sigma^2/(2M^2\omega^2)$. For $d=1$ the one loop expression for $Z_\phi(\omega,0)$ remains infrared divergent even for $\omega>0$. In any case, the negative one loop correction (\ref{A.13}) drives $Z_\phi(\omega,0)$ rapidly towards zero as $\omega$ decreases, making the use of renormalization group improvement mandatory. This has been done in \cite{CWQP}. 

Let us now turn to the evaluation of $Z_\phi(\omega,\vec q)$ for general $\vec q$ and $\omega$. By suitable momentum shifts eq. (\ref{A.2}) can be written as
\begin{eqnarray}\label{A.16}
\Delta\bar P_{21}&=&\frac{i\omega\lambda^2_\phi\bar\phi^2_0}{M}
\int\limits_{\vec p}
\left(\vec p\ ^2-2M\lambda_\phi\bar\phi^2_0+\frac{\vec q\ ^2}{4}\right)H,\nonumber\\
H&=&\int_{p_0}\det\nolimits^{-1}\left(p+\frac q2\right)\det\nolimits^{-1}\left(p-\frac q2\right)
\end{eqnarray}
where
\begin{eqnarray}\label{A.17}
\det\left(p\pm\frac q2\right)&=&B_\pm+\left(p_0\pm\frac i2 \omega\right)^2,\\
B_\pm&=&\frac{1}{4M^2}\left(\vec p\pm\frac{\vec q}{2}\right)^2
\left[\left(\vec p\pm\frac{\vec q}{2}\right)^2+\sigma\right].\nonumber
\end{eqnarray}
For the $p_0$-integration we need the poles which are situated at $p_{0,1}=+i\sqrt{B_+}-\frac i2\omega$ , $p_{0,2}=+i\sqrt{B_-}+\frac i2\omega$, $p_{0,3}=-i\sqrt{B_+}-\frac i2\omega~,~p_{0,4}=-i\sqrt{B_-}+\frac i2\omega$. For $|\omega|<2B_\pm$ one finds
\begin{eqnarray}\label{A.18}
H&=&i(r_1+r_2),\nonumber\\
ir_1&=&\left\{2\sqrt{B_+}\left(B_--(\sqrt{B_+}-\omega)^2\right)\right\}^{-1},\nonumber\\
ir_2&=&\left\{2\sqrt{B_-}\left(B_+-(\sqrt{B_-}+\omega)^2\right)\right\}^{-1}.
\end{eqnarray}
For $\vec q=0$ one recovers the formulae discussed above.

For $q=|\vec q|\to\infty$ we can neglect $\sigma$ such that
\be\label{A.19}
B_\pm=\frac{1}{4M^2}\left(\left(\vec p\pm\frac{\vec q}{2}\right)^2\right)^2.
\ee
We may evaluate eq. \eqref{A.18} for $\omega=0$
\ba\label{A.20}
H^{-1}&=&2\sqrt{B_+B_-(B_++B_-)}\nn\\
&=&\frac{1}{2\sqrt{2}M^3}
\left[\left(\vec p\ ^2+\frac{\vec q\ ^2}{4}\right)^2-
(\vec p\vec q)^2\right]\nn\\
&&\sqrt{\left(\vec p\ ^2+\frac{\vec q\ ^2}{4}\right)^2+
(\vec p\vec q)^2}.
\ea
This yields
\ba\label{A.21}
&&\Delta Z_\phi=2\sqrt{2}\lambda^2_\phi\bar\phi^2_0M^2\int_{\vec p}
\left({\vec p\ ^2+\frac{\vec q\ ^2}{4}}\right)\nn\\
&&\left[\left(\vec p\ ^2+\frac{\vec q\ ^2}{4}\right)^2-(\vec p\vec q)^2\right]^{-1}\nn\\
&&\left[\left(\vec p\ ^2+\frac{\vec q\ ^2}{4}\right)^2+(\vec p\vec q)^2\right]^{-1/2}.
\ea
The $\vec p$-integral is finite for $d<4$ and dominated by values $\vec p\ ^2\approx \vec q\ ^2/4$. As a result it is proportional $v_dq^{d-4}/(d-4)$ and we find an expression as in eq. \eqref{N13}. Solving the integral explicitely determines the coefficient $\gamma$ in this equation.

\section*{APPENDIX B: Fermion boson coupling}
\renewcommand{\theequation}{B.\arabic{equation}}
\label{fermionboson}
\setcounter{equation}{0}
In this appendix we briefly discuss the effects of fermion fluctuations on the bosonic occupation numbers in a coupled system as given by eq. \eqref{T1}. For this purpose the effective action has to be extended by a piece quadratic in the boson field $\bar \phi$ and by suitable boson interactions. We denote the inverse boson propagator by $\bar P_\phi=\bar G^{-1}$. We consider here an approximation where $\bar P_\phi$ is given by a classical part and a part $\Delta P_\phi$  induced by the fermion fluctuations
\be\label{B.A1}
\bar P_\phi=iq_0+\frac{\vec q\ ^2}{4M}-2\sigma+\Delta \bar P_\phi.
\ee
(The boson mass is twice the fermion mass $M$ and the chemical potential twice the fermion chemical potential. This corresponds to di-atom molecules.) This inverse propagator has been extensively studied in \cite{DGPW}. We are interested here in its value for analytically continued momenta $q_0=i\omega$ and at $T=0$. We want to show here that the fermion fluctuation contribution to the boson (or molecule) occupation number $n_M(\vec q)$ vanishes in absence of a gap $\Delta=\bar h_\phi\bar\phi_0=h_\phi\phi_0$ (for $T=0$). 

Using the shorthands 
\begin{equation}\label{B.A2}
\hat\omega=M(\omega+2\sigma),\hat q_0=2Mq'_0~,~\bar q_0=Mq_0
\end{equation}
one finds
\begin{eqnarray}\label{B.A2a}
\Delta\bar P_\phi(q_0+i\omega,\vec q)&=&-\frac{\bar h^2_\phi M}{\pi}\int_{\vec q\ '}\int d\hat q_0\nonumber\\
&&\Bigg\{\left[\left(\vec q\ '-\frac{\vec q}{2}\right)^2+i\bar q_0-\hat\omega+i\hat q_0\right]^{-1}\nonumber\\
&&\left.\left[\left(\vec q\ '+\frac{\vec q}{2}\right)^2+i\bar q_0-\hat\omega-i\hat q_0\right]^{-1}-\right.\nonumber\\
&&(\omega\to 0,\vec q\to 0,\bar q_0\to 0)\Bigg\}
\end{eqnarray}
The $\hat q_0$ integration has poles at
\begin{equation}\label{a2}
\hat q_0=i\left[\left(\vec q\ '-\frac{\vec q}{2}\right)^2-\hat\omega\right]-\bar q_0~,~
-i\left[\left(\vec q\ '+\frac{\vec q}{2}\right)^2-\hat\omega\right]+\bar q_0
\end{equation}
and can  be performed by closing the integral either in the upper or lower half plane
\begin{eqnarray}\label{a3}
\Delta\bar P_\phi=-\bar h^2_\phi M\int\limits_{\vec q\ '}
&&\Bigg\{\left[\vec q\ '^{2}+\frac{\vec q\ ^2}{4}-\hat\omega+i\bar q_0\right]^{-1}\nonumber\\
&&-\left[\vec q '^2-2M\sigma\right]^{-1}\Bigg\}.
\end{eqnarray}
We concentrate here on $d=3$ and $\sigma\leq 0$
\begin{eqnarray}\label{B.3A}
&&\Delta \bar P_\phi=\frac{\bar h^2_\phi M}{4\pi^2}\int\limits^\infty_{-\infty}dq'
\left\{\frac{A^2}{(q'-iA)(q'+iA)}+\right.\nonumber\\
&&\qquad \left.\frac{2M\sigma}{(q'-i\sqrt{-2M\sigma})(q'+i\sqrt{-2M\sigma})}\right\},\nonumber
\end{eqnarray}
with $A^2=\vec{q}\ ^2/4-\hat\omega+i\bar q_0$. We choose $A$ to be the root with positive real part. (This always exists for $\bar q_0\neq 0.$) The $q'$-integration yields
\begin{eqnarray}\label{a4}
\Delta\bar P_\phi=\frac{\bar h^2_\phi M^{3/2}}{4\pi}
\Big\{\left(\frac{\vec{q}\ ^2}{4M}+iq_0-\omega-2\sigma\right)^{1/2}_+
-\sqrt{-2\sigma}\Big\}\nn\\
\end{eqnarray}
where $(~)^{1/2}_+$ indicates the root with positive real part. We therefore obtain the propagator in the molecule phase $(\sigma<0)$
\begin{eqnarray}\label{b1}
&&G(q_0,\vec q)=\left[iq_0+\frac{\vec{q}\ ^2}{4M}\right.\\
&&\left.+\frac{\bar h^2_\phi M^{3/2}}{\pi}\left\{\left(iq_0+\frac{\vec{q}\ ^2}{4M}-2\sigma\right)^{1/2}_+-(-2\sigma)^{1/2}
\right\}\right]^{-1}.\nonumber
\end{eqnarray}

We want to show that the bosonic occupation number vanishes
\begin{equation}\label{b2}
n_{M}(\vec{q})=\int_{q_0}G(q_0,\vec q)=0.
\end{equation}
For this purpose we consider $\partial{n_{M}}(\vec{q})/\partial\bar h^2_\phi$. For this quantity we can close the integral on the lower half plane. The integral is analytic in the lower half plane, implying $\partial{n_{M}}(\vec{q}\ ^2)/\partial \bar h^2_\phi=0$. Since for $\bar h^2_\phi=0$ the theory is free we infer for all $\vec{q}\ ^2>0$ that $n_{M}(\vec q;~\bar h^2_\phi=0)=0$. Therefore $n_{M}(\vec{q})$ vanishes for all $\bar h^2_\phi$. 

For the atom phase one has $\sigma=0$ and the bosonic propagator becomes
\begin{eqnarray}\label{b3}
&&G(q_0,\vec q)=[iq_0+\frac{\vec{q}\ ^2}{4M}+\bar m^2_\phi\nonumber\\
&&\left.+\frac{\bar h^2_\phi M^{3/2}}{4\pi}
\left(iq_0+\frac{\vec{q}\ ^2}{4M}\right)^{1/2}_+\right]^{-1}
\end{eqnarray}
The argument showing $n_{M}(\vec{q})=0$ proceeds similar as above. 

In the limit \eqref{B.A1} the boson occupation number vanishes for $T=0$ as long as there is no gap. A non-vanishing fermion contribution to the boson numbers only arises from modifications of the fermion propagator in eq. \eqref{B.A2a} due to a non-vanishing gap $\Delta=\bar h_\phi\bar\phi_0$. Of course, the fermion fluctuation contribution to $n_M(\vec q)$ is nonzero for $T>0$ even in absence of a gap \cite{DW}. Furthermore, one has the contribution from boson fluctuations as discussed in the main part of this paper.

\end{document}